\pdfoutput=1
\documentclass[twocolumn,twocolappendix]{aastex63}
\usepackage{afterpage}
\usepackage{multirow}
\usepackage{listings}
\usepackage[tbtags]{amsmath}
\usepackage{hyperref}
\usepackage[normalem]{ulem} 

\newcommand{\change}[1]{{\bf \color{purple} #1}}

\definecolor{mypink1}{rgb}{0.858, 0.188, 0.478}
\definecolor{mypink2}{rgb}{0.99, 0.45, 0.75}
\definecolor{mygreen}{rgb}{0.0, 0.5, 0.1}

\defcitealias{arnaud2010}{A10}
\defcitealias{comis2011}{C11}
\defcitealias{marrone2012}{M12}
\defcitealias{planelles2017}{P17}
\defcitealias{vikhlinin2006b}{V06}

\begin{document}

\title{Pressure profiles and mass estimates using high-resolution Sunyaev-Zel'dovich effect observations of Zwicky 3146 with MUSTANG-2}
\author[0000-0001-5725-0359]{Charles E. Romero}
\affiliation{Department of Physics and Astronomy, University of Pennsylvania, 209 South 33rd Street, Philadelphia, PA, 19104, USA}
\author[0000-0001-6903-5074]{Jonathan Sievers}
\affiliation{Department of Physics, McGill University, 3600 University Street Montreal, QC H3A 2T8, Canada}
\author[0000-0002-3736-8058]{Vittorio Ghirardini}
\affiliation{Harvard-Smithsonian Center for Astrophysics, 60 Garden Street, Cambridge, MA 02138, USA}
\author{Simon Dicker} 
\affiliation{Department of Physics and Astronomy, University of Pennsylvania, 209 South 33rd Street, Philadelphia, PA, 19104, USA}
\author[0000-0002-1634-9886]{Simona Giacintucci}
\affiliation{U.S. Naval Research Laboratory, 4555 Overlook Avenue SW, Washington, DC 20375, USA}
\author[0000-0003-3816-5372]{Tony Mroczkowski}
\affiliation{ESO - European Southern Observatory, Karl-Schwarzschild-Str.\ 2, D-85748 Garching b.\ M\"unchen, Germany}
\author[0000-0002-8472-836X]{Brian S.\ Mason}
\affiliation{National Radio Astronomy Observatory, 520 Edgemont Rd., Charlottesville VA 22903, USA}
\author[0000-0003-0167-0981]{Craig Sarazin} 
\affiliation{Department of Astronomy, University of Virginia, P.O. Box 400325, Charlottesville, VA 22901, USA}
\author[0000-0002-3169-9761]{Mark Devlin}
\affiliation{Department of Physics and Astronomy, University of Pennsylvania, 209 South 33rd Street, Philadelphia, PA, 19104, USA}
\author[0000-0003-2754-9258]{Massimo Gaspari}
\altaffiliation{{\it Lyman Spitzer Jr.} Fellow}
\affiliation{Department of Astrophysical Sciences, Princeton University, 4 Ivy Lane, Princeton, NJ 08544, USA}
\author{Nicholas Battaglia}
\affiliation{Department of Astronomy, Cornell University, Space Sciences Building, Ithaca, New York 14853}
\author{Matthew Hilton}
\affiliation{Astrophysics and Cosmology Research Unit, University of KwaZulu-Natal, Westville Campus, Durban 4041, South Africa}
\author[0000-0002-7619-5399]{Esra Bulbul}
\affiliation{Max-Planck-Institute f\"ur extraterrestrische Physik (MPE), Gie{\ss}enbachstra{\ss}e 1, D-85748 Garching b.\ M\"unchen, Germany}
\affiliation{Harvard-Smithsonian Center for Astrophysics, 60 Garden Street, Cambridge, MA 02138, USA}
\author{Ian Lowe}
\affiliation{Department of Physics and Astronomy, University of Pennsylvania, 209 South 33rd Street, Philadelphia, PA, 19104, USA}
\author{Sara Stanchfield}
\affiliation{Department of Physics and Astronomy, University of Pennsylvania, 209 South 33rd Street, Philadelphia, PA, 19104, USA}

\shorttitle{A MUSTANG-2 SZ Study of Zw3146}
\shortauthors{Romero et al.}

\begin{abstract}
The galaxy cluster Zwicky 3146 is a sloshing cool core cluster at $z=0.291$ that in X-ray imaging does not appear to exhibit significant pressure substructure in the intracluster medium (ICM).
The published $M_{500}$ values range between $3.88^{+0.62}_{-0.58}$ to $22.50 \pm 7.58 \times 10^{14}$ M$_{\odot}$, where ICM-based estimates with reported errors $<20$\% suggest that we should expect to find a mass 
between $6.53^{+0.44}_{-0.44} \times 10^{14}$ M$_{\odot}$ (from Planck, with an $8.4\sigma$ detection) and $8.52^{+1.77}_{-1.47} \times 10^{14}$ M$_{\odot}$ (from ACT, with a $14\sigma$ detection). 
This broad range of masses is suggestive that there is ample room for improvement for all methods. Here, we investigate the ability to estimate the mass of Zwicky 3146 via the Sunyaev-Zel'dovich (SZ) effect with data taken at 90 GHz by MUSTANG-2 to a noise level better than $15\ \mu$K at the center, and a cluster detection of $104\sigma$. 
We derive a pressure profile from our SZ data which is in excellent agreement with that derived from X-ray data.
From our SZ-derived pressure profiles, we infer $M_{500}$ and $M_{2500}$ via three methods -- $Y$-$M$ scaling relations, the virial theorem, and  hydrostatic equilibrium -- where we employ X-ray constraints from \emph{XMM-Newton} on the electron density profile when assuming hydrostatic equilibrium.
Depending on the model and estimation method, our $M_{500}$ estimates range from $6.23 \pm 0.59$ to $10.6 \pm 0.95 \times 10^{14}$ M$_{\odot}$, where our estimate from hydrostatic equilibrium, is $8.29^{+1.93}_{-1.24}$ ($\pm 19.1$\% stat) ${}^{+0.74}_{-0.68}$ ($\pm 8.6$\% sys, calibration) $\times 10^{14}$ M$_{\odot}$. 
Our fiducial mass, derived from a $Y$-$M$ relation is 
$8.16^{+0.44}_{-0.54}$ ($\pm 5.5$\% stat) ${}^{+0.46}_{-0.43}$ ($\pm 5.5$\% sys, $Y$-$M$) ${}^{+0.59}_{-0.55}$ ($\pm 7.0$\% sys, cal.) $\times 10^{14}$ M$_{\odot}$.
We investigate the consistency of our mass estimates and potential for otherwise unaddressed systematic errors within the MUSTANG-2 data.
The inconsistencies among our mass estimates are not well explained by a potential systematic error in the MUSTANG-2 data.

\end{abstract}


\section{Introduction}
\label{sec:intro}
    
The spatial density of galaxy clusters as a function of mass and redshift can be used to constrain cosmological models (see \citealt{allen2011} and \citealt{Pratt2019} for reviews). 
However, the strength of these cosmological constraints is predominantly limited by the ability to accurately and precisely estimate masses \citep[e.g.][]{salvati2018} 
For galaxy cluster surveys using the Sunyaev-Zel'dovich (SZ) effect (\citealt{sunyaev1970,sunyaev1972}; see \citealt{Mroczkowski2019} for a recent review), mass estimates largely come from the $Y$-$M$ scaling relation 
\citep[e.g.][]{planck2014_XX,bleem2015,hilton2018}, where $Y$ is the integrated SZ signal. 
The precision of mass estimates from the $Y$-$M$ scaling relation suffer from the intrinsic scatter within these relations \citep[e.g.][]{salvati2018,sifon2013}, while the accuracy can suffer from systematic uncertainties and measurement biases, especially those in the data sample used to establish the scaling relation.

It is convenient to parameterize the strength of the thermal Sunyaev-Zel'dovich effect in terms of the Compton $y$ parameter, which is proportional to the thermal electron pressure integrated along the line of sight, $\ell$:
\begin{equation}
    y = \frac{\sigma_\textsc{T}}{m_e c^2} \int_0^\infty P_e d\ell,
    \label{eqn:Comptony}
\end{equation}
where $\sigma_\textsc{T}$ is the Thomson cross section, $m_e$ is the electron mass, $c$ speed of light, and $P_e$ is the thermal electron pressure.
The integrated $Y$ is thus a volumetric integral of the thermal electron pressure, usually computed within a cylinder or a sphere \citep[e.g.][]{motl2005,mroczkowski2009,arnaud2010}. We expect that when the gas is fully virialized its energy content is dominated, and thus approximated, as thermal energy. In this case, the thermal energy is directly related to gravitational potential energy 
\citep{kaiser1986,mroczkowski2011}, and thus there is a direct relation between $Y$ and $M$. 
Even in relaxed, virialized systems, we expect some non-thermal energy, and hence some non-thermal energy support \citep[e.g.][]{biffi16}.
So long as any non-thermal pressure support (non-thermal kinetic energy) evolves in a self-similar manner, an unbroken power law relation for $Y$-$M$ will hold. 

From SZ observations, there are three common methods of estimating the mass of a cluster: one employs a scaling relation, another employs hydrostatic equilibrium (HE), and the third employs the virial theorem (VT).
Masses derived using HE or VT generally omit non-thermal pressure support and, at least for HE, they will consequently be biased low with respect to the total mass \citep[e.g.][ and references therein]{miyatake2019}. The omission of non-thermal pressure support has generally been forced due to lack of access to this quantity. However, novel approaches such as pressure fluctuation analyses \citep{khatri2016}, use of high-resolution X-ray spectrometers \citep{lau2017}, and use of prior knowledge of gas fractions \citep{eckert2019} appear to be viable routes to quantifying non-thermal pressure support. 

In this paper, we analyze recently obtained deep, high resolution SZ data on a relatively nearby 
\citep[$z=0.291$,][]{allen1992}, relaxed, massive galaxy cluster, Zwicky 3146. 

Being a relaxed cluster of fairly circular shape on the sky means that assuming spherical symmetry should be sufficient, and thus it is an apt cluster to investigate the suitability of the above three mass estimates. A newly developed data processing method allows credible constraints on the pressure profile beyond our (radial) field of view and thereby has enabled us to constrain $M_{500}$ with MUSTANG-2 data alone. In addition to fitting pressure profiles, we quantify the residual signal after subtracting point sources and a spherical cluster model.

The layout of this paper is as follows. 
In Section~\ref{sec:Zw3146_overview}, we discuss the cluster we targeted for observations.
In Section~\ref{sec:obs} we discuss observations with MUSTANG-2 and review the data processing pipelines. 
In Section~\ref{sec:FittingMethod} we present our map-fitting procedures. In Section~\ref{sec:results} we present our results for the pressure profiles along with any additional components which had been fit. We discuss our results, largely focusing on best methods for mass estimation from SZ data in Section~\ref{sec:discussion} and conclude in Section~\ref{sec:conclusions}.

Throughout this paper, we adopt a concordance cosmology: $H_0 = 70$~km~s$^{-1}$~Mpc$^{-1}$, $\Omega_M = 0.3$, $\Omega_{\Lambda} = 0.7$. We define $h_{70} \equiv H_0$~(70 km~s$^{-1}$~Mpc$^{-1}$)$^{-1}$ and $h(z) \equiv H(z) H_0^{-1}$. At $z=0.291$, one arcsecond corresponds to 4.36~kpc. Uncertainties assume Gaussian distribution when presented with $\pm$ format, or when values are expressed as ${M}^{+U}_{-L}$, $M$ is the median, and $U$ and $L$ express the difference from the median to reach the 16$^{th}$ and 84$^{th}$ percentiles. We report literature values in the latter format, even if the original work provides results in the former format. 
For results from this work, if multiple uncertainties are presented, the first set is statistical, and the last is systematic due to flux calibration; if a third (middle) set is presented, this is the systematic error due to the $Y$-$M$ relation.

\section{The case of Zwicky 3146}
\label{sec:Zw3146_overview}
    
\begin{deluxetable*}{cccccc}   
\tabletypesize{\scriptsize}
\tablecolumns{6}
\tablecaption{Zwicky 3146 Mass Estimates \label{tbl:zwicky_masses}}
\tablehead{ 
  \colhead{$M_{2500}$} & \colhead{$M_{500}$} & \colhead{$M_{200}$} & \colhead{Facility} & \colhead{Cosmology} & \colhead{Estimation} \\
  \colhead{($10^{14} M_{\odot}$)} & \colhead{($10^{14} M_{\odot}$)} & \colhead{($10^{14} M_{\odot}$)} & \colhead{}
  & \colhead{$\Omega_M$,$\Omega{\Lambda}$,$h$} & \colhead{method} 
}
\startdata
    -- & $8.13$ & -- & ROSAT$^{1,2}$ & 0.3,0.7,0.75 & HE, isothermal $\beta$ \& NFW \\
        \vspace{0.5mm}
    $4.5$ & $10.8$ & -- & SuZIE$^{3}$ & 0.3,0.7,1.0 & HE, isothermal $\beta$ \& NFW \\
        \vspace{0.5mm} 
    $3.60^{+1.70}_{-1.70}$ & $8.65^{+4.09}_{-4.09}$ & $12.3^{+5.8}_{-5.8}$ & $^\ddagger$NOT$^{4,5}$ & 0.3,0.7,0.7 & Shear (WL) \& NFW \\ 
        \vspace{0.5mm} 
	-- & $10.81^{+5.90}_{-4.64}$ & -- & $^\ddagger$NOT$^{6}$ & 0.3,0.7,0.7 & Shear (WL) \& NFW \\
        \vspace{0.5mm}
	-- & $12.07^{+5.27}_{-5.27}$ & $18.72^{+8.18}_{-8.18}$ & $^\ddagger$NOT$^{5,6}$ & 0.3,0.7,0.7 & Shear (WL) \& NFW \\
        \vspace{0.5mm}
    $5.41^{+0.81}_{-0.81}$ & $22.50^{+7.58}_{-7.58}$ & -- & \emph{CXO}$^7$ & 0.3,0.7,0.7 & HE, $\beta$ \& RTM  \\
    -- & $^{\ast}20.8$ & $28.1^{+\infty}_{-16.3}$ & \emph{CXO}$^{8,9}$ & 0.3,0.7,0.7 & HE $\beta$ \& NFW \\
        \vspace{0.5mm}
	-- & $^{\ast}5.2$ & $^{\dagger}7.0^{+6.3}_{-2.3} \,^{+1.2}_{-0.6}$ & WFI$^{10}$ & 0.3,0.7,0.7 & Shear (WL) \& NFW \\
        \vspace{0.5mm}
	-- & $^{\ast}7.2$ & $9.7^{+16.3}_{-6.0}$ & \emph{XMM}$^{10}$ & 0.3,0.7,0.7 & HE $\beta$ \& NFW \\
        \vspace{0.5mm}
    $6.0^{+1.9}_{-1.6}$ & $^{\ast}14$ & -- & \emph{CXO}$^{11}$ & 0.3,0.7,0.7 & HE $\beta$ \& NFW \\
        \vspace{0.5mm}
	$3.6^{+0.2}_{-0.2}$ & $^{\ast}8.5$ & -- & \emph{CXO} and OVRO/BIMA$^{12}$ & 0.3,0.7,0.7 & HE \& isothermal $\beta$ \\
        \vspace{0.5mm}
	$3.3^{+0.5}_{-0.4}$ & $^{\ast}7.7$ & -- & OVRO/BIMA$^{12}$ & 0.3,0.7,0.7  & HE \& isothermal $\beta$ \\
        \vspace{0.5mm}
    -- & $6.72^{+0.44}_{-0.43}$ & $10.11^{+0.81}_{-0.77}$ & \emph{XMM}$^{13}$ & 0.3,0.7,0.7 & HE \& NFW \\
        \vspace{0.5mm}
    -- & $8.29^{+0.43}_{-0.42}$ & $13.55^{+1.53}_{-1.42}$ & \emph{XMM}$^{13}$ & 0.3,0.7,0.7 & HE, double $\beta$ \& NFW \\ 
        \vspace{0.5mm}
    -- & $7.85^{+0.23}_{-0.23}$ & -- & \emph{CXO}$^{14}$ & 0.3,0.7,0.7 & double $\beta$ and $Y_X$-$M^{15}$ \\
        \vspace{0.5mm}
    -- & $^{\ast}3.36$ & $4.54$ & SDSS$^{16}$ & 0.3,0.7,0.72 & $L_{\text{1 Mpc}}$-$r_{200}$ relation$^{17}$ \\
        \vspace{0.5mm} 
	$2.89^{+0.06}_{-0.06}$ & $6.82^{+0.14}_{-0.14}$ & -- & \emph{XMM}$^{18}$ & 0.3,0.7,0.7 & HE, modified $\beta$ and parametrized $T$ \\
        \vspace{0.5mm}
	$3.27^{+0.41}_{-0.41}$ & $7.83^{+0.96}_{-0.96}$ & -- & \emph{CXO}$^{18}$ & 0.3,0.7,0.7  & HE, modified $\beta$ and parametrized $T$\\
        \vspace{0.5mm}
	-- & $7.1$ & -- & \emph{CXO}$^{19}$ & 0.3,0.7,0.7 & HE $\beta$ \& NFW  \\
        \vspace{0.5mm}
    -- & $6.53^{+0.44}_{-0.44}$ & -- & Planck$^{20}$ & 0.3, 0.7, 0.7 & UPP, $Y$-$M^{21}$ \\
    $1.69^{+0.34}_{-0.37}$ & $3.88^{+0.62}_{-0.58}$ & $5.45^{+1.10}_{-0.98}$ & Subaru$^{22}$ & 0.3,0.7,? & Shear (WL \& NFW)  \\ 
    -- & $8.53^{+1.77}_{-1.47}$ & -- & ACTPol$^{23}$ & 0.3,0.7,0.7 &  UPP, $Y$-$M$ \\ 
    \enddata
\tablecomments{The Facility column refers to the primary instrument(s) used
to derive the respective mass. Some masses are not presented (at the tabulated density contrasts) in the original work, but are calculated in another work; in this case we list both references. $^{\ast}$If no $M_{500}$ is found in the literature, we use an average conversion from $M_{2500}$ or $M_{200}$ as described in the text. The quoted error bars (except $^\dagger$) are from as reported in the literature and vary with respect to the inclusion of systematic uncertainties. NFW refers to the \citet{navarro1997} profile; RTM refers to the \citet{rasia2004} profile; UPP refers to the Universal Pressure Profile \citep{arnaud2010}. $^\dagger$ The additional uncertainty ($^{+1.2}_{-0.6}$ ) reflects best fit values under different galaxy selections. $^{\ddagger}$NOT is the Nordic Optical Telescope. References: 
$^1$\citet{ettori1999}; $^2$\citet{mccarthy2003b}; $^{3}$ \citet{benson2004}; $^{4}$ \citet{dahle2006}; $^5$\citet{sereno2015a}; 
$^{6}$ \citet{pederson2007}; $^7$ \citet{morandi2007}; $^8$\citet{schmidt2007}; $^9$\citet{groener2016};  $^{10}$ \citet{kausch2007};
$^{11}$\citet{allen2008}; $^{12}$ \citet{bonamente2008}; $^{13}$\citet{ettori2011}; $^{14}$\citet{lancaster2011}; 
$^{15}$ \citep{vikhlinin2009a}; $^{16}$\citet{wen2013}; $^{17}$\citet{wen2012}; $^{18}$\citet{martino2014}; 
$^{19}$\citet{walker2014}; $^{20}$\citet{planck2016_XXVII}; $^{21}$\citet{planck2014_XX};
$^{22}$\citet{okabe2016}.}; $^{23}$\citet{hilton2018}
\end{deluxetable*}



Zwicky 3146 is cross-listed under several names including ACT-CL J1023.6+0411, BLOX J1023.6+0411.1, PSZ2 G239.43+47.95, RXC J1023.6+0411, and ZwCl 1021.0+0426.
Despite being in the Zwicky catalog \citep{zwicky1961}, Zwicky 3146 (alternatively listed as ZwCl 1021.0+0426) appears to have gone largely unscrutinized until \citealt{allen1992}, when it was detected in the ROSAT All Sky Survey (RASS) and followed up through optical spectroscopy from the Faint Object Spectrograph on the Isaac Newton Telescope. \citet{allen1992} found the BCG in Zwicky 3146 to be the most line-luminous in their sample. While line emission is common in cooling flow clusters, the observed line luminosities in Zwicky 3146 were found to be well above that expected from recombination of cooling IGM \citep{johnstone1987}. \citet{edge1994} estimate a pure cooling flow rate of 1250 M$_{\odot}$ yr$^{-1}$.
Subsequent reported rates vary between 300--1600 M$_{\odot}$~yr$^{-1}$ \citep{egami2006a,kausch2007,mcdonald2018}, which is on the upper end of the distribution of pure cooling flow rates. However, such theoretical rates are expected to be quenched down to 10\% via active galactic nucleus (AGN) feedback (e.g., \citealt{gaspari13}).
We may thus expect greater variability in the core due to feedback processes.

Zwicky 3146 is also remarkable for its $H_2$ mass, both in a cool and warm state \citep{egami2006b}. However, in the radio and hard X-rays, Zwicky 3146 is not marked by superlatives \citep{cooray1998a,nevalainen2004}. FIRST \citep{white1997} reveals a 2 mJy central radio source at 1.4~GHz (see Table~\ref{tbl:zwicky_radio}). 
\citet{giacintucci2014} used VLA data at 4.9 GHz and 8.5 GHz to image the center of Zwicky 3146, finding two sources at each frequency, but only 4.9 GHz shows a radio minihalo. \citet{kale2015} extended this radio analysis, adding of 610 MHz GMRT data. At 610 MHz, the minihalo shows the same extent and is roughly 7 times brighter than at 4.9 GHz, implying a spectral index of $\alpha \sim 0.9$, which is slightly shallower than typical minihalo spectral indices \citep[$\alpha \sim 1.2$-$1.3$, as reported in][]{giacintucci2014}.

Simply stated, the interesting part of Zwicky 3146 appears to be its core, where the strong cooling flow and radio minihalo are of particular interest. There is some expectation that, in galaxy clusters in general, these two features will have some correlation \citep{mcnamara2007,bravi2016}. \citet{forman2002} identify three edges (or fronts) in \emph{Chandra} data within the central 30\arcsec\ which they propose may be due to sloshing. Averaged over the entire cluster, \citet{hashimoto2007} find a minor/major axis ratio of 0.85 using data from the \emph{Chandra} archive. \citet{weismann2013} define Zwicky 3146 as being a ``regular'' cluster, which they define by the lack of substructure in two different Gaussian smoothing kernels ($\sigma = 4$\arcsec\ and 8\arcsec). Given that $\sigma = 4$\arcsec\ corresponds closely to the MUSTANG-2 resolution, we should expect minimal substructure in our maps. 

The first mass estimate for Zwicky 3146 that we are aware of comes from X-ray studies \citep{rasia2004,ettori1999}, and followed shortly thereafter by an SZ study \citep{mccarthy2003b}. When deriving masses from the ICM, nearly all studies have used a spherical beta ($\beta$) model \citep{cavaliere1978} to fit the gas density profile. Often an additional assumption (if only for simplicity or lack of constraint) is that the ICM is isothermal. While this is sufficient to calculate a mass under HE, many papers have added the constraint that the total mass profile be fit by a NFW \citep{navarro1997} profile:
\begin{equation}
    \rho(r) = \frac{\rho_0}{(r/R_s)(1+r/R_s)^2},
    \label{eqn:NFW}
\end{equation}
where $\rho_0$ and $R_s$ are the matter density normalization and scaling radius, respectively. Weak lensing studies have calculated surface mass profiles from shear and subsequently fit those profiles to a NFW profile. It is striking that for the same, or ostensibly very similar, assumptions, and especially use of the same dataset, the mass estimates vary by almost an order of magnitude; see Table~\ref{tbl:zwicky_masses}. 

Not all literature masses are provided at $M_{500}$. In this case, we adopt an average conversion from $M_{2500}$ or $M_{200}$. It is generally found that $0.6 < R_{500}/R_{200} < 0.7$, and that $R_{500}/R_{200} \approx 0.66$ \citep[e.g.][]{shimizu2003,shaw2008,battaglia2012a}. If we take $\xi = R_{500}/R_{200}$, then $M_{500} / M_{200} = (\xi)^3 *500 / 200 = 0.74$ for $\xi = 2/3$. As in \citet{arnaud2005}, we use $R_{2500} = 0.44 R_{500}$, or $M_{500} = 2.35 M_{2500}$. 
In Table~\ref{tbl:zwicky_masses}, we do not estimate errors bars for these extrapolated values.

Calculating an expected value from the literature is non-trivial given that the datasets are not all independent (many make use of the same underlying data) and systematic errors are often not reported. We expect that our MUSTANG-2 derived mass should agree more strongly with other ICM-based masses (i.e. those derived from X-ray or SZ observations).  Of the ICM-based masses with reported errors which are $< 20$\%, the highest estimate is 
$8.52^{+1.77}_{-1.47} \times 10^{14}$ M$_{\odot}$. 
and the lowest estimate (even without the uncertainty criterion) is
$6.53^{+0.44}_{-0.44} \times 10^{14}$ M$_{\odot}$. 

The mass estimates (in units of $10^{14}$ M$_{\odot}$) not based on the ICM span from $3.88^{+0.66}_{-0.58}$ to $12.07^{+5.27}_{-5.27}$. Restricting the estimates to only those with reported error bars $< 20$\%, we arrive at only one estimate: $3.88^{+0.66}_{-0.58}$, from weak lensing \citep{okabe2016}. The tension between this estimate and the lowest one from the ICM (that from Planck) is at $3.3\sigma$. 

For reference, with our adopted cosmology at $z=0.291$, $M_{500} = 8 \times 10^{14}$ M$_{\odot}$  corresponds to $R_{500} = 1280$ kpc or 293\arcsec~and $M_{2500} = 3.5 \times 10^{14}$ M$_{\odot}$ corresponds to $R_{2500} = 567$ kpc or 130\arcsec.

\section{Observations and Data Reduction}
\label{sec:obs}

\subsection{MUSTANG-2 observations}
\label{sec:M2_obs}
   
MUSTANG-2 is a 215-element array of feedhorn-coupled TES bolometers \citep{dicker2014a}. Observing at 90 GHz on the 100-meter Robert C. Byrd Green Bank Telescope (GBT), MUSTANG-2 achieves a resolution of 10\arcsec~and has an instantaneous field of view (FOV) of 4.\arcmin25.  MUSTANG-2 is the successor to the MUSTANG instrument \citep{dicker2008}, which had 64 detectors with the same resolution (determined by the telescope optics and coupling), but with only a 42\arcsec\ FOV. The increased FOV of MUSTANG-2 enables us to recover cluster-sized scales required for the work presented here.

Our observational techniques and data reduction are largely the same as with MUSTANG \citep[see][]{romero2015a,romero2017}. We briefly review them here. Absolute flux calibrations are preferentially based on the Solar System objects Mars, Uranus, Jupiter, Saturn, and Ceres. We additionally use ALMA grid calibrators \citep{fomalont2014,vanKempen2014}, where the latest observations can be accessed through \url{https://almascience.eso.org/sc/}. At least one of the above flux calibrators was observed once per night. To track the telescope pointing and gain, we observe a point source every 30-50 minutes.

\begin{figure}
  \begin{center}
     \includegraphics[width=0.45\textwidth]{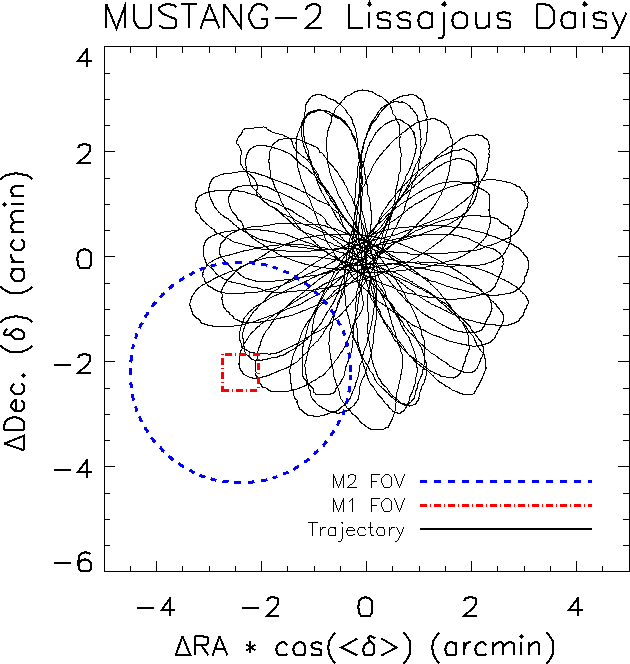}
     \includegraphics[width=0.45\textwidth]{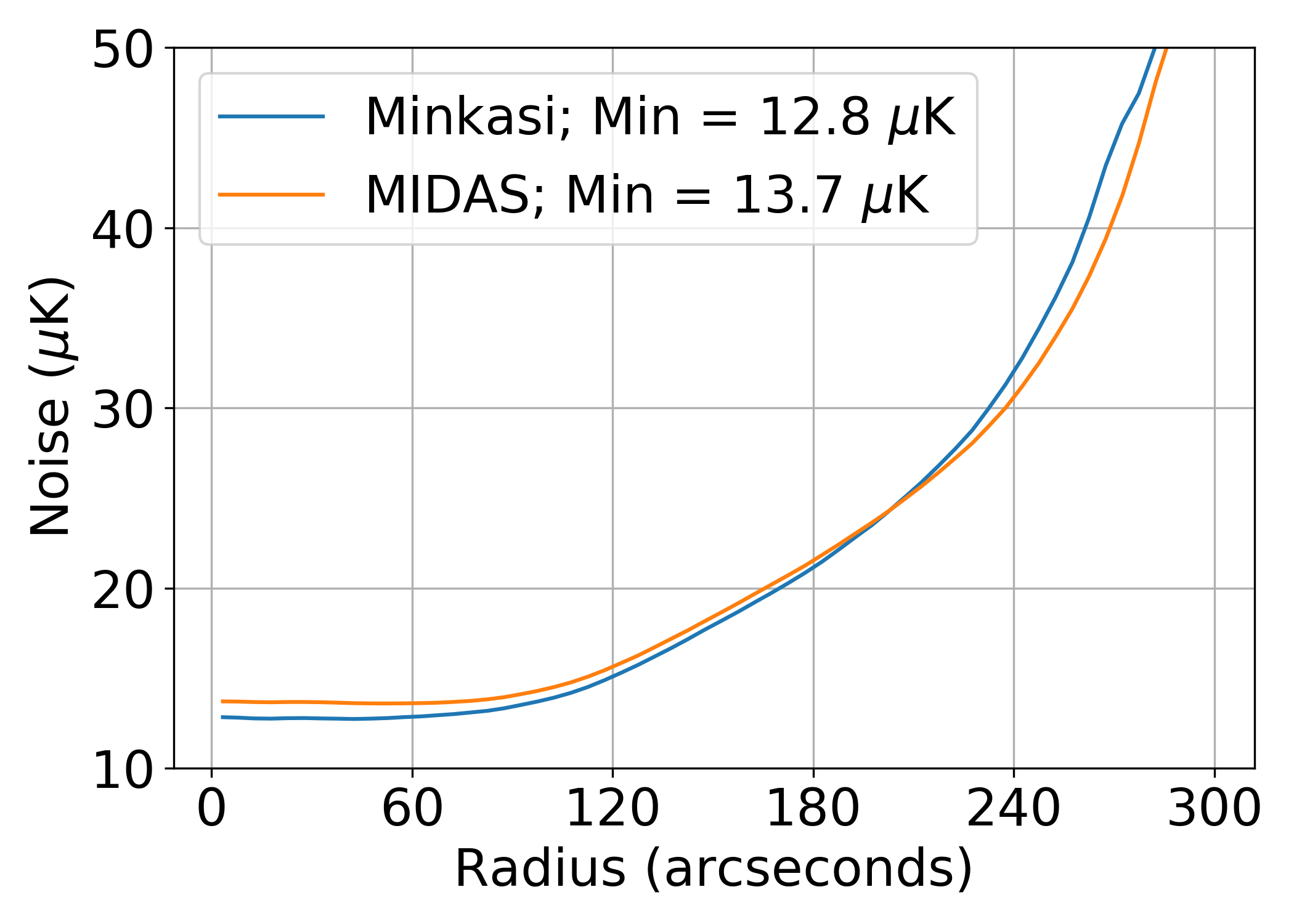}
  \end{center}
  \caption{Top: Example scan pattern for MUSTANG-2 (M2), which uses the same scan strategy as MUSTANG (M1), scanning at $0.6\arcmin$ per second, on average. The FOV of the two instruments is plotted for comparison. Bottom: Noise (RMS) profiles from the two pipelines. Of note is that the profile is relatively flat in the inner 2\arcmin. }
  \label{fig:m2_scan_pattern}
\end{figure}

To observe the target galaxy cluster, we employ Lissajous daisy scans, typically with a $3\arcmin$ radius (see Figure~\ref{fig:m2_scan_pattern}). In the first observing session on Zwicky 3146, we tested scan radii of $2.5$, $3$, and $3.5$ arcmin and found that a $3\arcmin$ radius gave the best mapping speed. Our larger scans employ faster scan speeds, which are generally preferred as it shifts the sky signal to higher frequencies, beyond the 1/f noise of the atmosphere or readout. However, to ensure proper pointing, the jerk (third derivative of position, $\dddot{\mathbf{x}}$) of the GBT places limits on our scan speed, given the scan pattern. Each of the three scan patterns give a roughly uniform coverage in the central $\sim 2\arcmin$. Zwicky 3146 was observed under project ID AGBT18A\_175 on the nights of 2018 Feb 01, 2018 Mar 18, 2018 Mar 24, 2018 Mar 31, 2018 Dec 12, 2018 Dec 28, and 2019 Jan 11 with a total on-source integration time of 22.7 hours. Excising bad scans, our final maps incorporate 21.4 hours of data and have a RMS noise of $15~\mu$K (when smoothed to beam resolution) within the central $2\arcmin$ radius; a more detailed RMS profile is shown in Figure~\ref{fig:m2_scan_pattern}.

\subsection{MUSTANG-2 data reduction}

  We have developed two methods for the processing of MUSTANG-2 data. The first method, called the MUSTANG IDL Data Analysis System (MIDAS), builds off the custom IDL pipeline used with MUSTANG \citep{romero2015a,mason2019}; the second method is a maximum likelihood approach (Minkasi). Both techniques include the same initial quality checks (within IDL) performed on the raw data, which we refer to as time ordered data (TOD) from each of the responsive detectors. This paper focuses on the results obtained through the Minkasi pipeline. A comparison of the performance between the two approaches is in Appendix~\ref{sec:appendix_midas_minkasi}; here we offer a summary distinction between the two pipelines.
  
  The primary product produced from MIDAS is a (2D) map, where the TODs have been filtered to subtract atmospheric and electronic signal, thus leaving the sky signal. The methods employed to do this subtract some sky signal as well, acting as a high-pass filter and limiting the scales recovered to those within the FOV. In contrast, Minkasi is based on fitting a sky model to the TODs. This sky model can be a map, or for this work, a set of concentric annuli. In both cases, Minkasi does not impose a high-pass filter as MIDAS does. In other words, we can recover scales beyond our FOV. With respect to the concentric annuli, another advantage is that we recover a complete covariance matrix for the annuli, whereas we would need to estimate the covariance matrices for our maps, either in MIDAS or Minkasi.

  \subsubsection{Minkasi pipeline}
  
    The initial steps of preparing data for processing through Minkasi are as follows:
  (1) A pixel mask is defined based on the responsivity of the detectors from the instrument setup at the beginning of the run; unresponsive detectors are masked out.
  (2) Gain and opacity corrections are applied to our data.
  (3) A noise template is constructed as the common mode across all detectors. This template and a high ($\sim20$) order polynomial are simultaneously fit to and subtracted from the TODs
  (4) The cleaned TODs are checked for glitches and a small portion of the TOD (from just before to just after each glitch) is flagged. Detector weights are assigned based on the RMS of the corresponding TOD.
  (5) Rather than make a map, we save the uncleaned, but calibrated TOD (before step 3) along with detector weight information (as calculated at step 4).

  From here, these saved TODs are passed to the main Minkasi pipeline. The Minkasi pipeline requires some model be proposed. For mapping, each pixel has a corresponding model TOD. The stack of all pixel model TODs is denoted under canonical notation as $\mathbf{A}$. The noise (covariance) ($\mathbf{N}$) of our data is modelled in Fourier space. In particular, within the Fourier domain, we perform a singular value decomposition (SVD) across detectors per scan. Our noise model is taken as a smoothed version of the SVD-rotated power spectra. Because the number of pixels (and thus models) is so large, exact solutions cannot be obtained directly. Rather, we use a preconditioned conjugate gradient (PCG) descent to obtain our best-fit maps. We find sufficient convergence with 30 to 60 steps in the PCG. This process within Minkasi can be iterated, whereby the previous map (as a TOD) is subtracted when modelling the noise. The entirety of the (original) data, $\mathbf{d}$, is still used when calculating $\mathbf{A} \mathbf{N}^{-1} \mathbf{d}$. 

\begin{figure*}
  \begin{center}
    \includegraphics[width=0.31\textwidth]{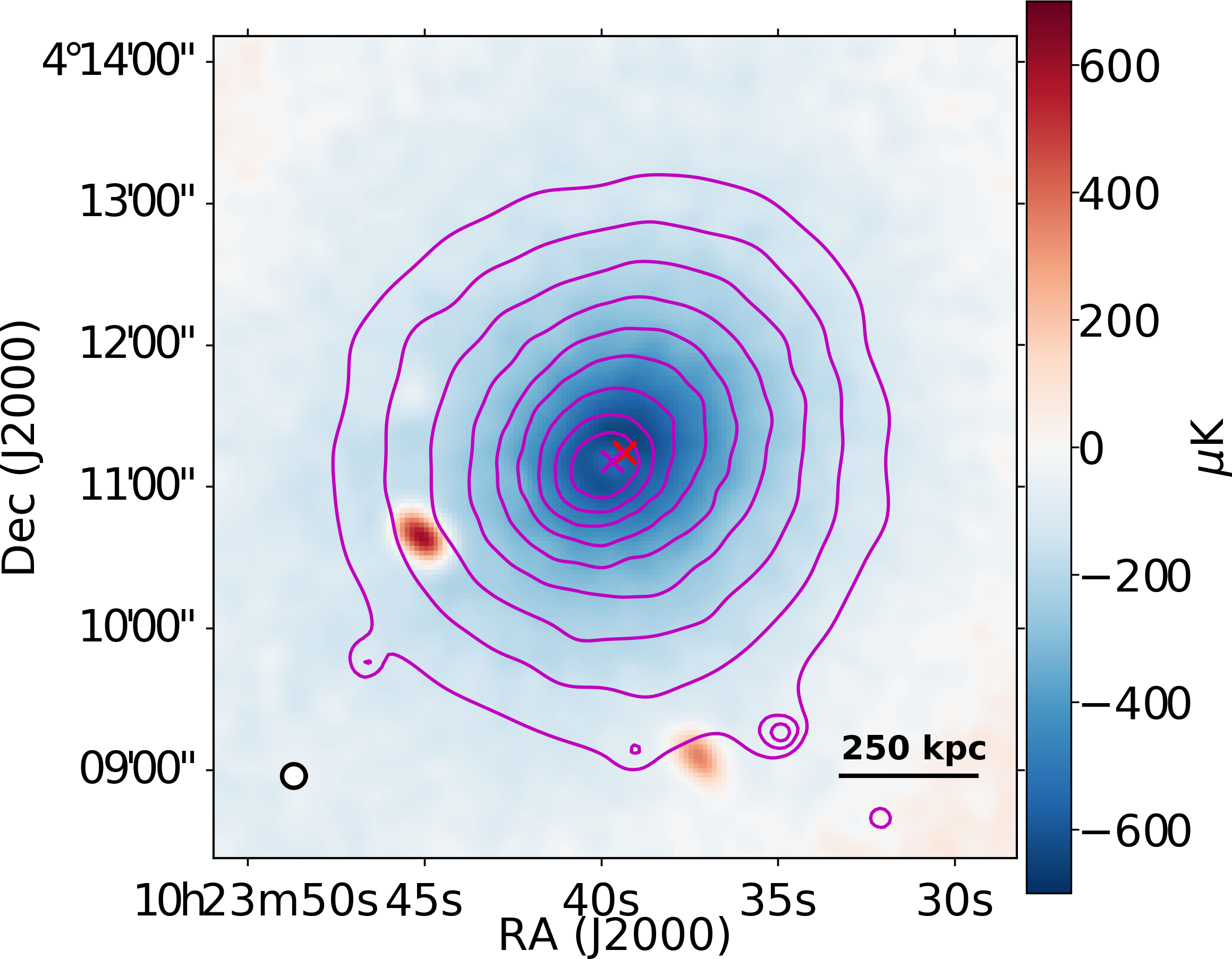} 
    \includegraphics[width=0.31\textwidth]{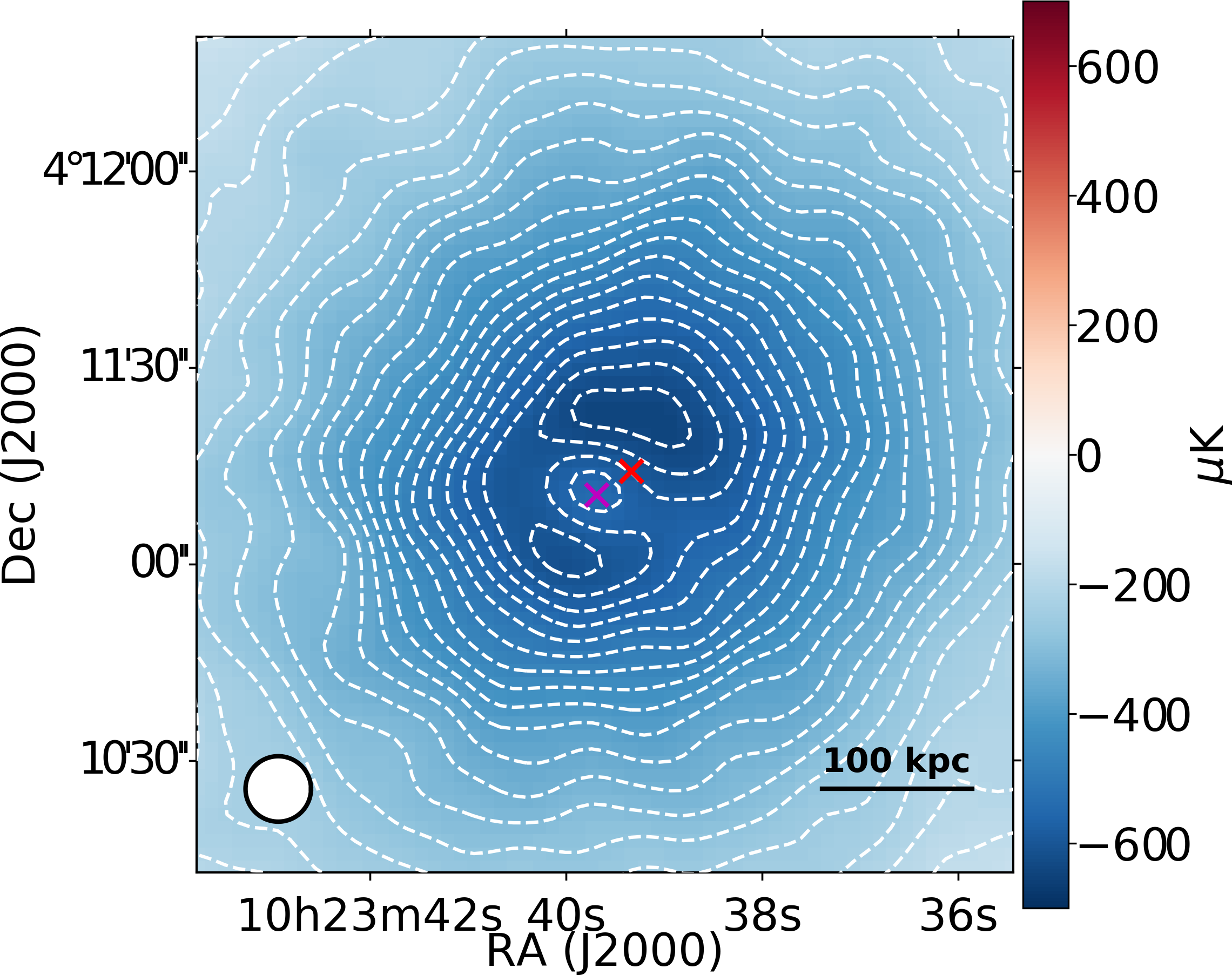}
    \includegraphics[width=0.31\textwidth]{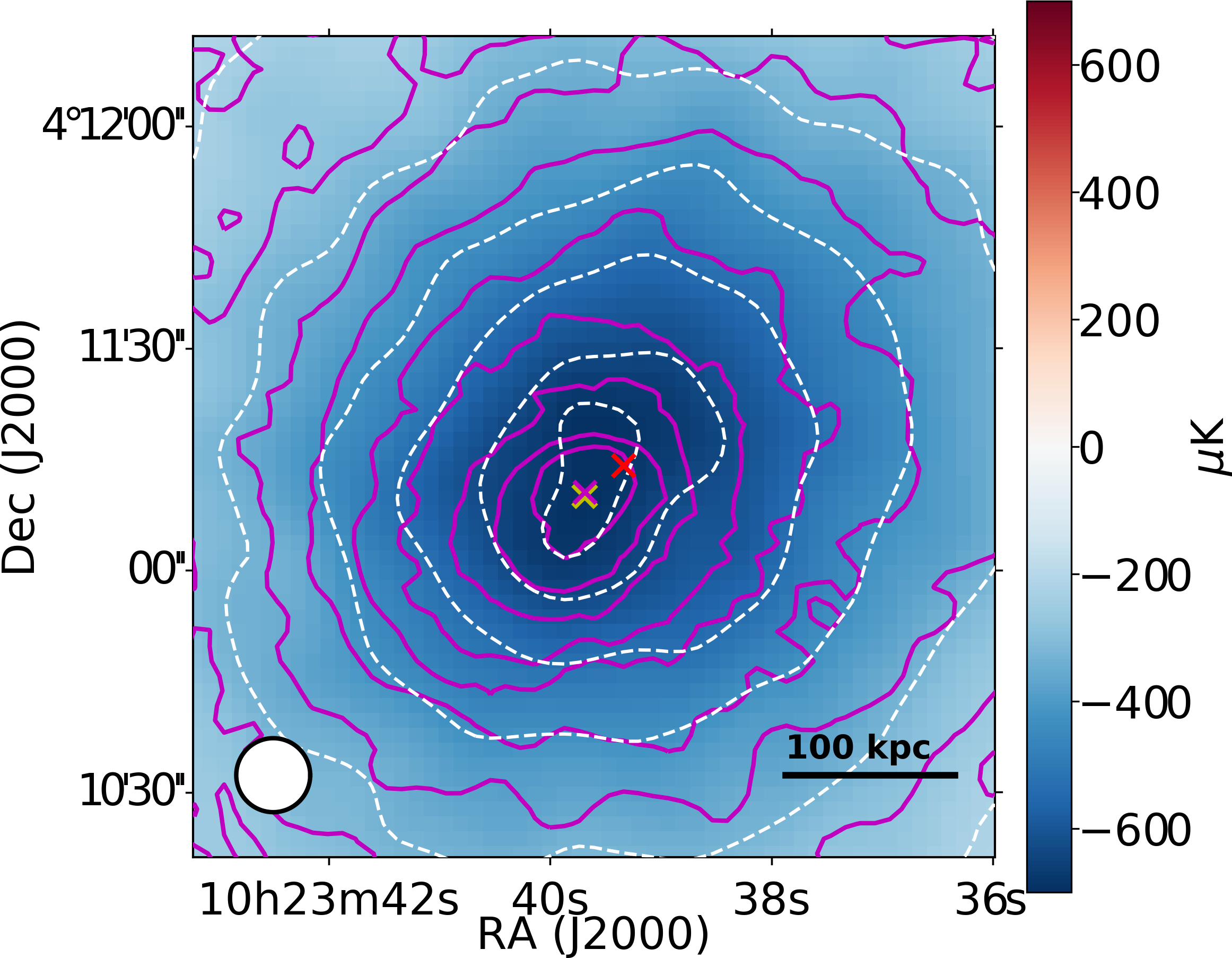}
  \end{center}
    \label{fig:m2_CARMA_zw3146}
  \caption{MUSTANG-2 images of Zw3146.  Left: Minkasi map, zoomed out, with smoothed X-ray surface brightness contours from \emph{XMM} in magenta. Middle: Minkasi map, zoomed in, with significance contours (every
  $2\sigma$) in white. Right: point-source subtracted
  Minkasi map (Section~\ref{sec:pp_fitting})
  with contours (white) at [-54, -50, -42, -34, -26, -18]$\sigma$ and X-ray (\emph{Chandra}) surface brightness contours overlaid in magenta.
  The red cross denotes the SZ centroid; the magenta cross denotes the X-ray (\emph{XMM}) centroid.}
\end{figure*}

  \subsubsection{Surface brightness profiles}
  
   Given an assumed cluster geometry, we opt to fit a surface brightness profile directly to the MUSTANG-2 timestreams within the Minkasi pipeline. The advantage of this approach (over using the map produced with Minkasi) is that we can explicitly solve for the profile and thus recover an accurate covariance matrix.
  To do this our (gridding) matrix, which transforms between map and time space, is the same as before. Rather than fit for individual pixel values, we fit for annuli. The annuli may either assume a constant (uniform) brightness within it, or assume a slope dictated by a previous iteration. A model for each annulus is constructed, convolved with the MUSTANG-2 beam (in map space) and converted to a timestream model. Thus, the resulting fits represent a deconvolved surface brightness profile. In the deconvolved surface brightness profile, we find a peak decrement of $1002 \pm 70$ $\mu$K, or $2.93 \times 10^{-4}$ in Compton y. When smoothed by the MUSTANG-2 beam, the peak decrement is $741 \pm 13$ $\mu$K, or $2.17 \times 10^{-4}$ in Compton $y$.

\section{Fitting pressure profiles}
\label{sec:FittingMethod}
    
\subsection{Compact source removal}

    There are six radio sources identified by FIRST \citep{becker1994} which correspond to sources found in the MUSTANG-2 maps. Four of the sources appear as two pairs, which from MUSTANG-2 data alone appear non-point-like. All six of these sources are well modelled as point sources. The centroids and FWHM are found in an initial round of fitting. The amplitudes are then fit simultaneously with the surface brightness profiles (Section~\ref{sec:M2_obs}). As discussed in Section~\ref{sec:resids}, we do find evidence for two additional point sources.
    At the redshift of Zwicky 3146, the conversion between the reported MUSTANG-2 flux densities (in Jy) and integrated Compton $Y$ (Mpc$^{2}$) is -0.00086 $h_{70}^{-2/3}$. 
    In the MUSTANG-2 data, the total flux density from all point sources is 3 mJy, which equivalent to $-2.6 \times 10^{-6} h_{70}^{-2/3}~\rm Mpc^2$, which is less than $\lesssim2.5$\% of the total $Y_{sph}(R_{500})$ that we find for this cluster.

\subsection{Pressure profile fitting}
\label{sec:pp_fitting}

The relation between the projected thermal SZ signal and a pressure profile lies in the Compton $y$ parameter, given by Equation~\ref{eqn:Comptony}. Using the notation in \citet{carlstrom2002} the change in temperature of the cosmic microwave background (CMB) due to the tSZ is given by:
\begin{equation}
    \frac{\Delta T_\textsc{CMB}}{T_\textsc{CMB}} =
 f(x)  y \, ,
\end{equation}
where $x = h \nu / k_\textsc{B} T_\textsc{CMB}$ is the scaled frequency, $h$ is the Planck constant, $\nu$ is the frequency, $k_\textsc{B}$ is the Boltzmann constant, and $T_\textsc{CMB}$ is the temperature of the CMB. The function $f(x)$ then governs the spectral distortion of the CMB. We correct for relativistic corrections; thus $f(x)$ is actually $f(x,T_e)$ as given by \citet{itoh1998}. As $f(x,T_e)$ is only a weak function of temperature for typical cluster temperatures ($k_B T_e < 15$~keV) and observations at 90~GHz, we assume $k_B T_e = 7$ keV, which is within the spread of temperatures reported in the literature (references cited in Table~\ref{tbl:zwicky_masses}). Additionally, our maps are calibrated to brightness temperature (i.e.\ a Rayleigh-Jeans brightness temperature, $T_\textsc{B}$). The conversion is comes from the derivative of the Planck function, and at 90 GHz, $\Delta T_\textsc{CMB}/T_\textsc{B} = 1.23$ \citep[see e.g.][]{Finkbeiner1999,Mroczkowski2019}.

In our non-parametric (NP) pressure profile model, we assume a power-law distribution of pressure within radial bins. We define twelve bins with twelve logarithmically spaced radii between 5\arcsec~and 5\arcmin. The innermost bin spans from 0 to 7\arcsec.25, while the slope and normalization are determined by the pressure at $r = 5$\arcsec~and $r=7$\arcsec.25. The outermost bin spans from 207\arcsec~to infinity with the slope and normalization determined by the pressure at $r = 207$\arcsec~and $r=300$\arcsec. All other bins span between neighboring logarithmically spaced radii (edges). We also directly fit the generalized NFW pressure profile \citep{nagai2007}:
\begin{equation}
    P(r) = \frac{P_0 P_{500}}{(C_{500} r / R_{500})^{\gamma}(1 + (C_{500} r / R_{500})^{\alpha})^{(\beta - \alpha)/\gamma}},
    \label{eqn:gnfw}
\end{equation}
where $P_0$, $C_{500}$, $\alpha$, $\beta$, and $\gamma$ were established as free parameters, and
\begin{equation}
    P_{500} = 1.45 \times 10^{-11} \left(\frac{M_{500}}{10^{15} h^{-1} M_{\odot}}\right)^{2/3} E(z)^{8/3},
\end{equation}
and $E(z)^2 = \Omega_{\rm M} (1+z)^3 + \Omega_{\Lambda}$. As many of the gNFW parameters are degenerate, interpreting physical meaning from the fitted values is more difficult when all parameters are allowed to vary. Thus, we choose to fix $\alpha$, $\beta$, and $\gamma$ to their respective values found in
\citealt{arnaud2010} (hereafter, A10):
1.05, 5.41, and 0.31; we refer to the gNFW profile with these restrictions as an A10 profile. The Compton $y$ profile for the A10 (pressure) profile is computed via numerical integration, with line-of-sight bounds at $\pm 5 R_{500}$, for the fiducial $M_{500} = 8 \times 10^{14}$ M$_{\odot}$. We find that our mass results are quite independent of the integration constraint (out to $5 R_{500}$); they vary by at most 1.1\% between adopting a fiducial mass $M_{500} = 16 \times 10^{14}$ M$_{\odot}$ and $M_{500} = 4 \times 10^{14}$ M$_{\odot}$.

The A10 pressure profile that we adopt here is slightly more general than the universal pressure profile (UPP) presented in \citetalias{arnaud2010}, where $c_{500}$ is fixed, and $P_{500}$ and $R_{500}$ are functions of mass ($M_{500}$). In Section~\ref{sec:pp_recovery}, we compare our NP and A10 fitted pressure profiles and discuss the impact on mass estimation in Section~\ref{sec:our_masses}.

 When fitting to the surface brightness profile found with Minkasi, the Compton $y$ profile is converted to a $T_\textsc{B}$ profile and matched to the bin normalizations reported by Minkasi. As Minkasi reports a full covariance matrix associated with the bin normalizations, this is used when determining the likelihood of each model. In both cases, the fitting is done with the MCMC package {\tt emcee} \citep{foreman2013}.

\begin{figure}
  \begin{center}
    \includegraphics[width=0.45\textwidth]{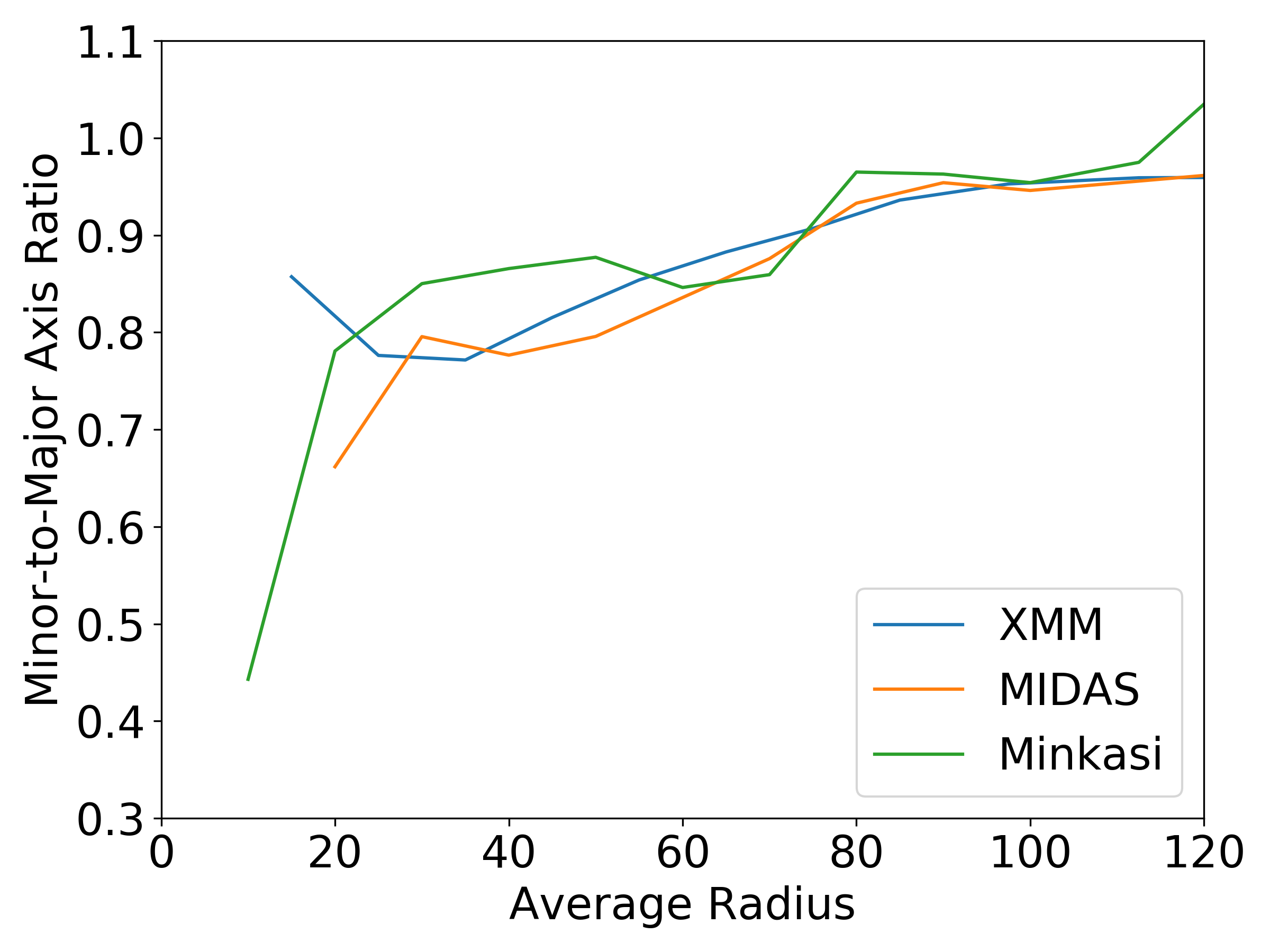}
  \end{center}
  \caption{Minor-to-major axis ratios (for axes in the plane of the sky) as determined via our two map-making procedures, with a comparison to the
  X-ray (\emph{XMM-Newton})
  image.}\label{fig:mink_rings}
\end{figure}

The choice of adopting spherical symmetry is motivated by (1) the ease of interpretation of results and comparison to the literature, and (2) our data suggest that a globally ellipsoidal model is not necessary. In Figure~\ref{fig:mink_rings}, we calculate axis ratios based on isophotes at several radii. While the center exhibits high ellipticity, beyond an arcminute, the cluster isophotes are close to circular. That is, an ellipse which fits the center will not be appropriate for the majority of the cluster-centric radii.

\subsection{Mass derivations}
\label{sec:mass_derivations}

    We investigate mass estimates via three avenues: (1) employing a $Y$-$M$ relation, (2) employing spherical hydrostatic equilibrium (HE) and (3) employing the virial theorem (VT) as in \citet{mroczkowski2011}. The first and third methods can be done with SZ data alone, while the second avenue requires knowledge of the gas density (or electron density), obtained here from analysis of \emph{XMM-Newton} data. 

    \subsubsection{via $Y$-$M$ relation}

    There are two flavors of integrated $Y$ quantities: $Y_{cyl} = Y_\textsc{SZ} D_\textsc{A}^2 $ \citepalias{arnaud2010} and $Y_{sph}$. The two are calculated as follows:
    \begin{align}
        Y_\textsc{SZ}(\theta) &= \frac{\sigma_\textsc{T}}{m_e c^2} \int_0^{\theta} \left[ \int_{-5 R_{500}}^{5 R_{500}} P_e(\ell) d\ell) \right] 2 \pi \theta d \theta \\
        Y_{sph}(R) &= \frac{\sigma_\textsc{T}}{m_e c^2} \int_0^R 4 \pi P_e(r) r^2 dr,
    \end{align}
    where
    \change{$r$}
    is the physical radius, $\theta$ is the angular (projected) radius, and for $Y_{cyl}$, $\ell$ is the distance along the line of sight. $Y_{cyl}$ is often reported in units of Mpc$^2$, rather than steradians. We opt to use the $Y_{sph}$ relation as it is easily integrated analytically.

    As a $Y$-$M$ relation only applies to a specific radius, we adopt $Y$-$M$ relations at $R_{500}$ and $R_{2500}$; at each radius, we calculate the mass via three different relations. At $R_{500}$, we adopt the relations in \citetalias{arnaud2010}, \citet[][hereafter M12]{marrone2012} , and \citet[][hereafter P17]{planelles2017}. At $R_{2500}$, we adopt the relations in \citet[][hereafter C11]{comis2011}, \citetalias{marrone2012} and \citetalias{planelles2017}. At each step in the MCMC pressure profile fits, we compute $Y_{sph}(R)$ and derive a self-consistent $R_{500}$ (see Appendix~\ref{sec:appendix_self_consistency}) for each $Y$-$M$ relation. Systematic uncertainties on $Y$-$M$ are calculated using the uncertainties on the $Y$-$M$ parameters reported the respective paper. If covariances are listed, they are used; otherwise, the uncertainties are assumed to be independent.

    \subsubsection{via hydrostatic equilibrium}

    Under hydrostatic equilibrium, pressure and gravity are related by:
    \begin{equation}
        \nabla{P} = - \rho_g \nabla{\phi},
    \end{equation}
    where $P$ is the total gas pressure, $\rho_g$ is the gas mass density, and $\phi$ is the gravitational potential \citep[see, e.g.][]{sarazin1988}. If we assume a spherically symmetric, non-rotating ICM with homogeneous metallicity and electron-ion equipartition, we can rewrite this as: 
    \begin{equation}
        \frac{dP_e}{dr} = - \mu m_p n_e G \frac{M_{\textsc{HE}}(<r)}{r^2},
        \label{eqn:hse1}
    \end{equation}
    where $\mu$ is the mean molecular weight, $m_p$ is the proton mass, $n_e$ is the electron (count) density, $G$ is the gravitiational constant, and $M_\textsc{HE}$ is the total (dark matter + baryonic) mass enclosed within radius $r$.  We adopt $\mu = 0.61$ \citep[e.g][]{eckert2019}.
    
    In practice, we rewrite Equation~\ref{eqn:hse1} as follows:
    \begin{equation}
        M_{\text{HE}} = -\frac{d \ln{P_e}}{d \ln{r}} \frac{P_e}{n_e} \frac{r}{\mu m_p G}
        \label{eqn:hse2}
    \end{equation}
    
    From SZ data alone, we lack $n_e$. If we assume a temperature profile (isothermal or, perhaps a Vikhlinin profile \citep[][hereafter V06]{vikhlinin2006b}), we can calculate $n_e$ from the SZ-derived $P_e$. However, a temperature normalization is still necessary. For Zwicky 3146, many temperatures are published in the literature. Given an intial mass ($M_{500}$) estimate, one could calculate $T_X$ from its scaling relation with $M$. In this work, we make use of an electron density profile from \emph{XMM} \citep[following][]{ghirardini2018}. Mass profiles are calculated as a function of radius and a self-consistent $R_{500}$ is calculated as shown in Appendix~\ref{sec:appendix_self_consistency} for each step in the MCMC.

    \subsubsection{via virial equilibrium}

    One can also assume that the ICM is in virial equilibrium, which derives from statistical mechanics and relates the total kinetic energy, $E_{kin}$ to the gravitational potential energy, $U_g$:
    \begin{equation}
        2 E_{kin}(R) = - U_g(R).
    \end{equation}
    \citet{mroczkowski2011} take the kinetic energy to be the thermal energy $E_{kin} = E_{th}$, while the erratum considers an external surface pressure term (\citealt{mroczkowskierratum2012}; see e.g.\ \citealt{kippenhahn2012} for a general discussion). This results in the virial relation in the form:
    $2 E_{th} - 3P(R)V = -U_{g}(R)$.  Here, $R$ is the outer radius of the system, and $V$ is the total volume.
    Assuming an NFW matter profile, they derive an expression which includes a definite integral. Here, we express this relation with the integral solved:
    \begin{equation}
    \begin{split}
        \frac{\mu_e}{16 \pi^2 G \mu f_{\text{gas}}}
        \left[3 \frac{m_e c^2}{\sigma_\textsc{T}} Y_{sph}(R) - 4 \pi R^3 P_e(R) \right] = \\
        \rho_0^2 R_s^5 \left[ \frac{R/R_S - \ln{(1 + R/R_s)}}{1 + R/R_S} - \frac{1}{2(1 + R_s/R)^2}\right],
    \end{split}
    \label{eqn:vir_mas}
    \end{equation}
    where $\mu_e$ is the electron mean molecular mass (mean molecular mass per electron), $f_{\text{gas}}$ is the gas fraction, $\rho_0$ and $R_s$ are the matter density normalization and scaling radius, respectively, in the NFW profile (Equation~\ref{eqn:NFW}). As in \citet{mroczkowski2011}, we take $\mu_e = 1.17$ and $f_{\text{gas}} = 0.13$. We note that $\mu_e = 1.17$ is consistent with $\mu = 0.61$ \citep{eckert2019}.
    Because this approach requires fitting the left hand side of Equation~\ref{eqn:vir_mas} to its right hand side, we do this at the end of the MCMC to use appropriate error bars. We exclude the points within the central 100 kpc, as is often done in X-ray analyses in cool core clusters; additionally, as discussed in Section~\ref{sec:results}, our constraints in the central 100 kpc are prone to degeneracies with the central point sources. Error bars on the NFW parameters $\rho_0$ and $1/R_s$ are computed within the least square fitting algorithm \lstinline[language=python]!scipy.opt.curve!. Subsequently, $M_{500}$ is calculated self-consistently from the NFW mass curve (as for the hydrostatic mass) over 1000 iterations with the covariance matrix on $\rho_0$ and $1/R_s$ as returned by the fitting routine. The parameters $\rho_0$ and $1/R_s$ are fit rather than $\rho_0$ and $R_s$ as the former pair have a roughly linear covariance.

\section{Results}
\label{sec:results}
    
\subsection{Pressure Profiles}
\label{sec:pres_profs}

\begin{figure*}
  \begin{center}
     \includegraphics[width=0.8\textwidth,clip=true, trim=2mm 2mm 2mm 3mm]{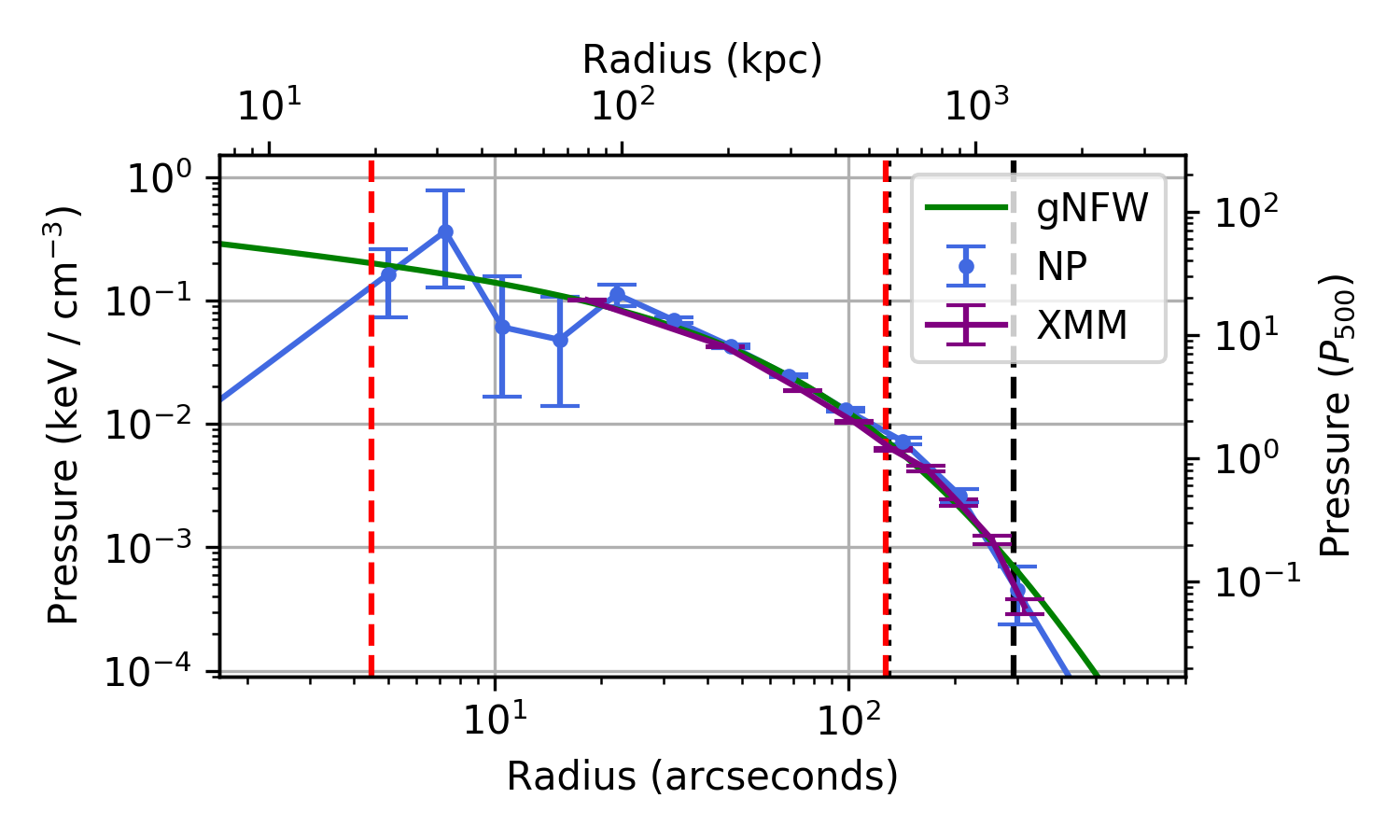}
  \end{center}
  \caption{
  Two fitted pressure profile models for Zwicky 3146, as well as the pressure profile as determined from \emph{XMM-Newton}. The vertical red dashed lines are the HWHM and radial FOV for MUSTANG-2; the vertical black dashed line is $R_{500}$ for $M_{500} = 8\times 10^{14} M_{\odot}$ and the vertical black dotted line (just to the right of the right red dashed line)
  is $R_{2500}$ for $M_{2500} = 3.5\times 10^{14} M_{\odot}$.
  }
  \label{fig:zw3146_pressure_profiles}
\end{figure*}

We expect our best pressure profile constraints over radial scales ranging from the half width, half max (HWHM) of the beam out to the radial FOV (i.e. from 5\arcsec\ to 126\arcsec\ for MUSTANG-2), where the constraints beyond the (radial) FOV are often quite limited for ground-based single dish SZ instruments \citep[e.g.][]{adam2014,czakon2015,romero2018,ruppin2018,sayers2018,dimascolo2019}.  Moreover, we may expect the tightest constraints around the geometric mean of the HWHM and radial FOV \citep{romero2015a}. As seen in Figure~\ref{fig:zw3146_pressure_profiles}, MUSTANG-2 produces tight constraints ($< 15$\%) between $30\arcsec < r < 240\arcsec$, with the tightest fractional constraint occurring in our bin just beyond 1\arcmin. The constraints in the center ($r < 30\arcsec$) are poor owing to the presence of two known radio sources, one of which was fit for, while the other we investigate in Section~\ref{sec:resids}. Additionally, our four innermost pressure profile bins sample the pressure profile in annuli whose widths are less than a beamwidth. Finally, ellipticity and low-level substructure may yet play into variations in our the pressure of our central most bins; we find that centroid choice (between the X-ray or SZ centroid) has a negligible impact on the pressure profile.

\subsection{Mass Estimates}
\label{sec:masses}

For each pressure profile we fit, we estimate the cluster mass via the three methods presented in Section~\ref{sec:mass_derivations}; our $M_{500}$ and $M_{2500}$ estimates are shown in Table~\ref{tbl:m2_mass_Minkasi}. We find that at $R_{500}$, $M_{\text{VT}}$ yields the largest mass estimates for each pressure profile and $M_{\text{HE}}$ tends to be larger than the $Y$-$M$ mass estimates. Additionally, the gNFW masses (both $M_{500}$ and $M_{2500}$) tend to be less than the respective NP masses; the exceptions are $M_{VT}$ at both overdensities and $M_{HE}$ at $M_{2500}$.

The agreement or disagreement of $M_{\text{HE}}$ with the other mass estimates also highlights a key difference between the methods of estimation: we calculate $M_{\text{HE}}$ at individual radii, which does not make use of knowledge of the pressure profile across all radii. However, the $Y$-$M$ mass makes use of $Y$, a cumulative sum of spherically integrated pressure within $R_{500}$ (or $R_{2500}$). Similarly, $M_{\text{VT}}$ is fit across all pressure bins at radii larger than 100 kpc. 

We note that $M_{\text{HE}}$ is higher than $Y$-$M$ mass estimates, where the latter, for all but ($Y$-$M$)$_{\text{A10}}$ and ($Y$-$M$)$_{\text{C11}}$ are taken as $M_{tot}$, the total gravitating mass (from weak lensing or numerical simulations). We expect $M_{\text{HE}}$ to be lower due to the so-called hydrostatic mass bias, which itself encompasses several biases. The primary bias is not actually the hydrostatic equilibrium assumption, but rather that the pressure gradient used only accounts for thermal pressure. Thus, the lack of account for non-thermal pressure under-estimates the mass by 10\% to 30\% \citep[see also][]{biffi16,khatri2016,hurier2018,ettori19}. 

\begin{deluxetable*}{l c c c c c c}
\tabletypesize{\scriptsize}
\tablecolumns{7}
\tablewidth{0pt}
\tablecaption{MUSTANG-2 Mass Estimates \label{tbl:m2_mass_Minkasi}}
\tablehead{\colhead{$M_{\Delta}$} & \colhead{Model} & \colhead{($Y$-$M$)$_1$} & \colhead{($Y$-$M$)$_2$} & \colhead{($Y$-$M$)$_3$} & \colhead{$M_{\textsc{HE}}$} & \colhead{$M_{\textsc{VT}}$} \\
 & \colhead{} & \colhead{($10^{14} M_{\odot}$)} & \colhead{($10^{14} M_{\odot}$)} & \colhead{($10^{14} M_{\odot}$)} & \colhead{($10^{14} M_{\odot}$)} & \colhead{($10^{14} M_{\odot}$)}}
\startdata
        & & & & & & \\
    \multirow{3}{*}{$M_{500}$} & NP & $8.16^{+0.44}_{-0.54} {}^{+0.46}_{-0.43} {}^{+0.59}_{-0.55}$ & $6.53^{+0.24}_{-0.24} {}^{+0.14}_{-0.14} {}^{+0.36}_{-0.34}$
                            & $7.85^{+0.49}_{-0.52} {}^{+0.10}_{-0.10} {}^{+0.61}_{-0.56}$ & $8.29^{+1.93}_{-1.24} {}^{+0.74}_{-0.68}$ & $9.05^{+0.56}_{-0.67} {}^{+0.74}_{-0.68}$ \\
        & & & & & & \\
                             & gNFW & $7.70^{+0.17}_{-0.17} {}^{+0.42}_{-0.39} {}^{+0.56}_{-0.52}$ & $6.23^{+0.10}_{-0.10} {}^{+0.16}_{-0.15} {}^{+0.34}_{-0.32}$
                             & $7.48^{+0.19}_{-0.17} {}^{+0.10}_{-0.09} {}^{+0.58}_{-0.54}$   & $8.33^{+0.47}_{-0.45} {}^{+1.00}_{-0.91}$ & $10.6^{+0.10}_{-0.10} {}^{+0.90}_{-0.80}$ \\
        & & & & & & \\
        \multirow{3}{*}{$M_{2500}$} & NP & $3.68^{+0.41}_{-0.33} {}^{+0.85}_{-0.69} {}^{+0.46}_{-0.42}$ & $3.23^{+0.25}_{-0.22} {}^{+1.39}_{-0.97} {}^{+0.34}_{-0.31}$
                             & $2.95^{+0.27}_{-0.28} {}^{+0.06}_{-0.06} {}^{+0.33}_{-0.31}$ & $3.62^{+2.76}_{-0.99} {}^{+4.09}_{-1.04}$ & $3.36^{+0.09}_{-0.11} {}^{+0.33}_{-0.30}$ \\
        & & & & & & \\
                            & gNFW & $3.50^{+0.13}_{-0.13} {}^{+0.81}_{-0.66} {}^{+0.43}_{-0.40}$ & $3.10^{+0.09}_{-0.09} {}^{+1.28}_{-0.90} {}^{+0.32}_{-0.30}$
                            & $2.83^{+0.10}_{-0.09} {}^{+0.06}_{-0.06} {}^{+0.31}_{-0.29}$ & $4.78^{+0.21}_{-0.20} {}^{+0.93}_{-1.54}$ & $3.96^{+0.02}_{-0.01} {}^{+0.39}_{-0.35}$ \\
         & & & & & & \\
\enddata
\tablecomments{ Mass estimates from MUSTANG-2 (and electron density profiles from \emph{XMM}) for $M_{\text{HE}}$. $M_{\text{VT}}$ is not the virial mass as is classically defined (with respect to $R_{vir}$), but rather the mass within $R_{\Delta}$ using the Virial theorem \citep{mroczkowski2011}. The error bars on the $Y$-$M$ mass are, in order, the statistical error, systematic error from the $Y$-$M$ relation itself, and the systematic error due to calibration uncertainty. The error bars on the other mass estimates are the statistical and systematic error due to calibration uncertainty. The $Y$-$M$ relations for $M_{500}$ for the subscripts 1, 2, and 3 are \citetalias{arnaud2010,marrone2012,planelles2017}, respectively; for $M_{2500}$ the relations come from \citetalias{comis2011,marrone2012,planelles2017}, respectively.}
\end{deluxetable*}


\subsection{Residuals}
\label{sec:resids}

\begin{figure}
  \begin{center}
     \includegraphics[width=0.45\textwidth]{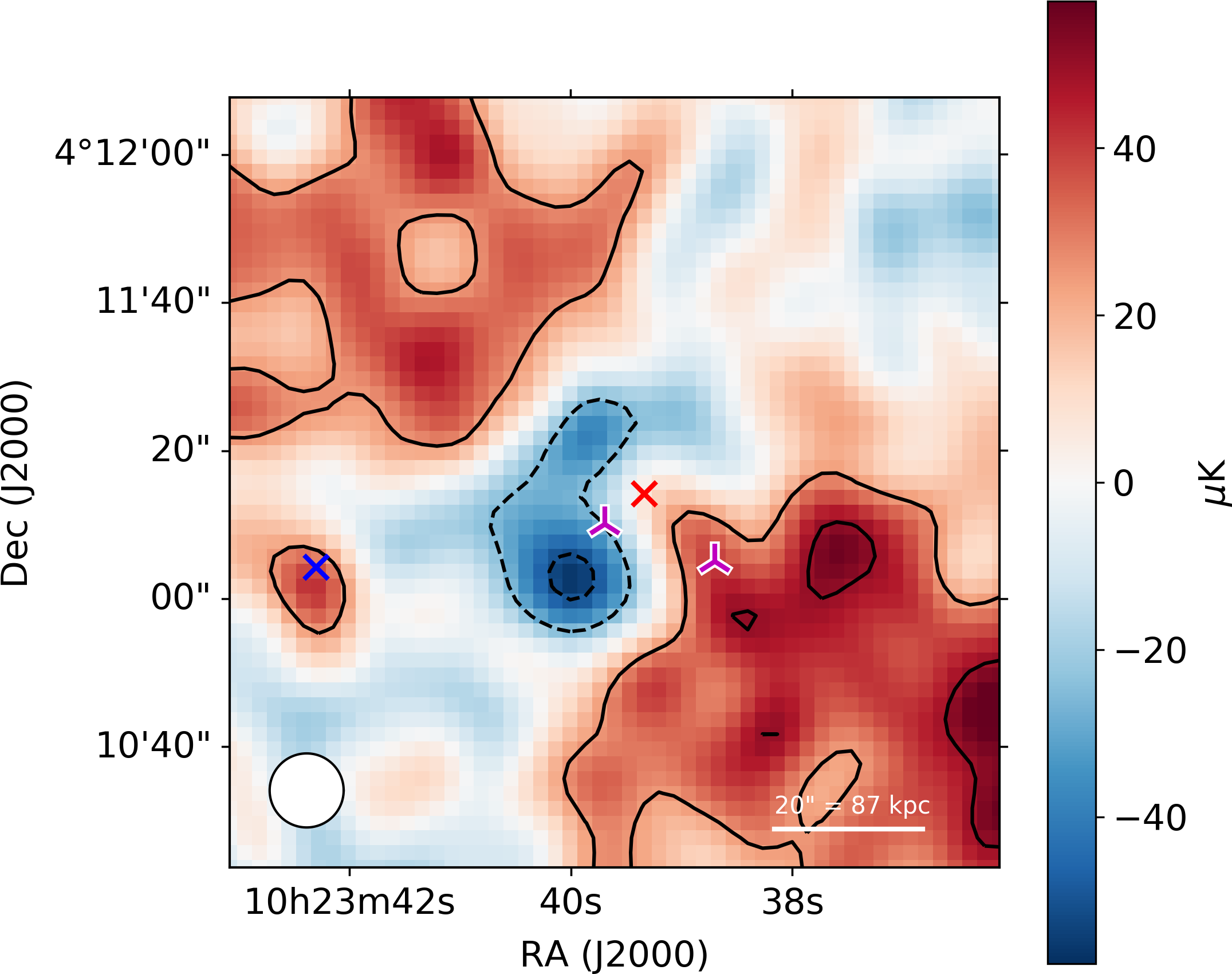}
  \end{center}
  \caption{The residual map (in $\mu$K) of the Minkasi map with the SZ centroid marked by a red cross, the radio sources marked by tri\_up (purple on white), and the Herschel 500 micron centroid by a blue cross; the beam is shown in the bottom left.
  The black contours are every $2\sigma$, excluding 0 ($\sigma = 12.8 \mu$K).}
  \label{fig:zw3146_resid1}
\end{figure}

An underlying quadropole is observed in our residual map (Figure~\ref{fig:zw3146_resid1}. Modelling the ICM with the same centroid and an ellipsoidal ICM distribution results in very similar residual decrements and removes this quadrupole. Beyond ellipticity in the core of the cluster, the only observed SZ substructure is also in the core of the cluster.  
The corresponding integrated Y ($Y_{cyl}$) due to the substructure (decrement $< -25 \mu$K) is $3.0 \times 10^{-7} h_{70}^{2/3}$ Mpc$^2$, whereas we find $Y_{cyl}(R_{500}) = 7.3 \times 10^{-5}h_{70}^{2/3}$ Mpc$^2$ for our non-parametric fit; thus substructure accounts for only 0.5\% of the SZ signal within $\theta_{500}$, and is not a significant source of scatter in the mass estimates from $Y$-$M$ relations for this cluster. 
    
As noted before, the central pressure profile and the central point source amplitude are degenerate. In addition to the central radio source, a second, nearby radio point source is observed at 4.9 and 8.5 GHz \citep{giacintucci2014}.
We find that these two sources contribute to a mild increase in the uncertainties in the inner two pressure profile bins. 

\begin{figure}
  \begin{center}
     \includegraphics[width=0.45\textwidth]{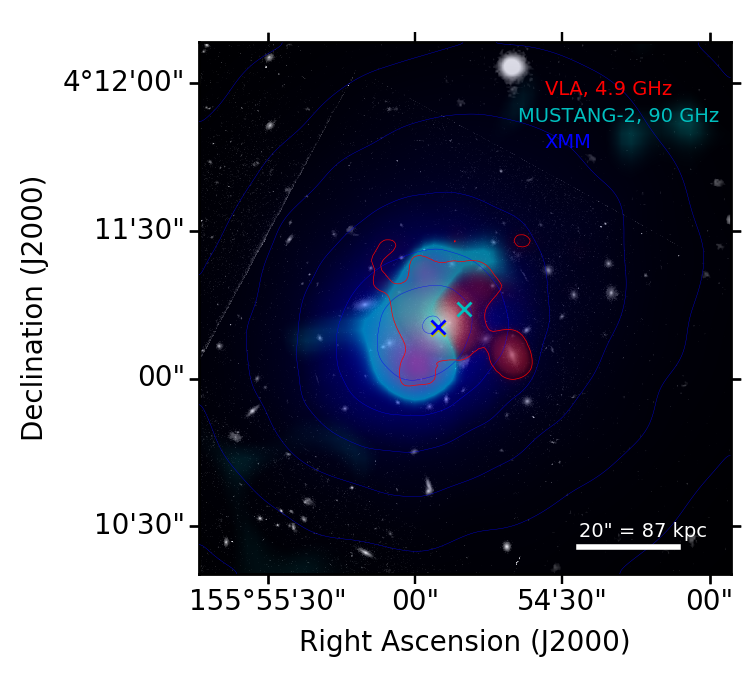}
  \end{center}
  \caption{HST image with MUSTANG-2 residual (cyan), X-ray (\emph{XMM}, blue), and radio (4.5 Ghz, red) overlays. The cyan (blue) cross is the X-ray (SZ) centroid.}
  \label{fig:zw3146_resid2}
\end{figure}

The spatial coincidence of our residual SZ decrement and the radio minihalo (Figure~\ref{fig:zw3146_resid2} may be suggestive of an underlying link. Equally interesting would be if there is a link between the SZ residual and the noted sloshing \citep{forman2002}.
    
\subsection{SZ/X-ray products}
\label{sec:add_prod}

 Given the SZ and X-ray data in hand, we extend our profile analyses to other thermodynamic quantities, namely temperature, and entropy. Given the better performance of Minkasi results (NP and gNFW), we do not include the MIDAS pipeline results in these analyses.
 
 \begin{figure}
  \begin{center}
     \includegraphics[width=0.45\textwidth]{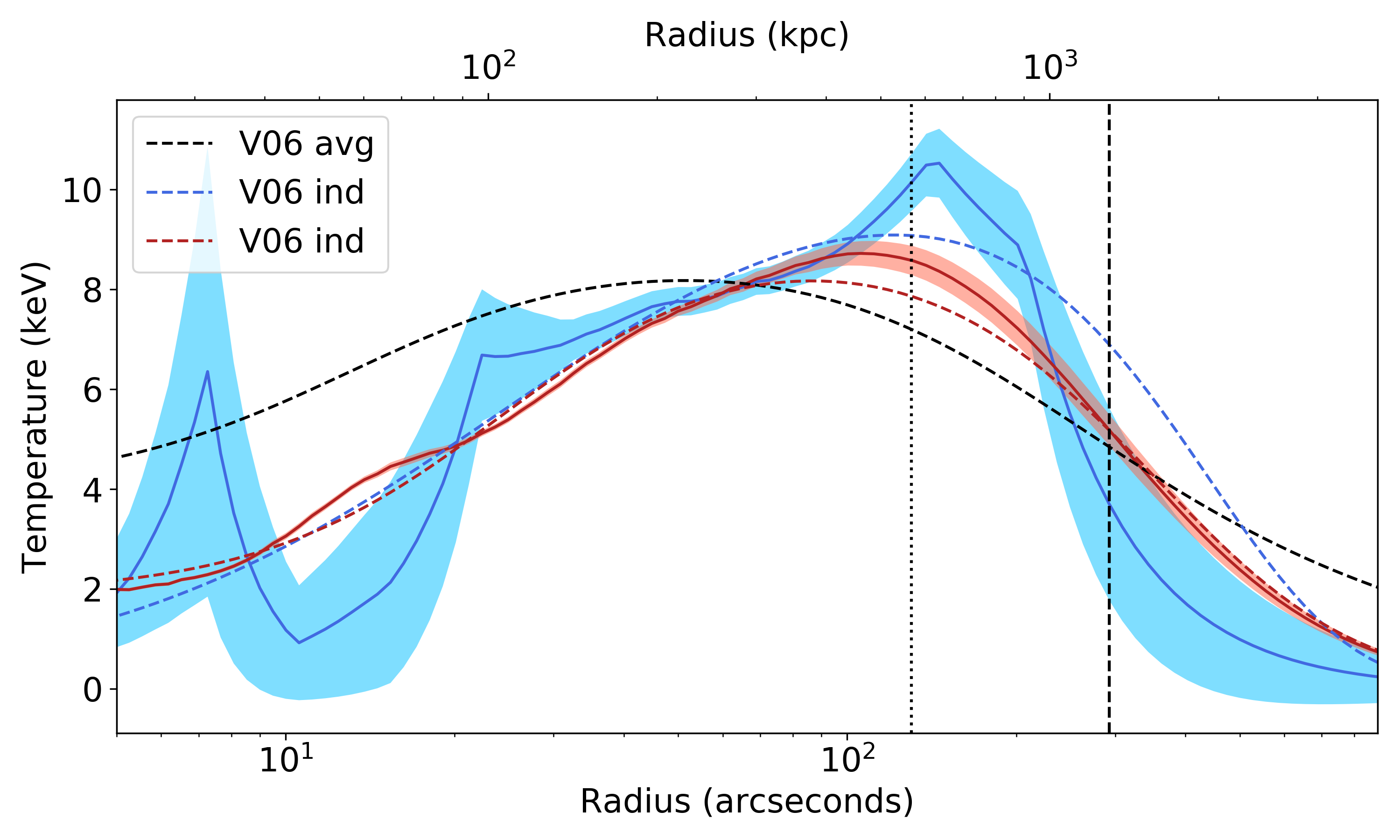}
  \end{center}
  \caption{Temperatures as inferred from the SZ (Minkasi pipeline) pressure and the X-ray-derived electron density. The red curve and uncertainty band is from the gNFW pressure profile and the blue curve is from the NP pressure profile. The black dashed temperature curve is the approximate average temperature profile found in \citetalias{vikhlinin2006b} (their Eq. 8), with $T_{mg}$ fitted. The blue and red dashed curves are the fitted individual Vikhlinin profiles (Eq.\ 6 in \citetalias{vikhlinin2006b}).
  The dotted and dashed vertical lines are as in Figure~\ref{fig:zw3146_pressure_profiles}.
  }
  \label{fig:zw3146_temperature}
\end{figure}

\subsubsection{Temperature}

We compute our temperature as
\begin{equation}
    k_B T_e = P_e / n_e,
\end{equation}
where $P_e$ is determined from our SZ data, and $n_e$ comes from X-ray data. We use $T$ in keV as shorthand for $k_B T_e$.

In order to assess how our temperature profile compares with others in the literature, we fit our results to two parameterizations from \citetalias{vikhlinin2006b} (hereafter, V06). The first (Equation~\ref{eqn:v06}) is a general parameterization, which should accommodate variations of shape between cluster temperature profiles. The second parameterization (Equation~\ref{eqn:v06avg}) was found as an approximation to the average temperature profile across the cluster sample in \citetalias{vikhlinin2006b}; it has a fixed shape, with only the normalization being allowed to vary. Therefore, comparison to this second curve gives some indication of how ``average" Zwicky 3146 is, with respect to the sample in \citetalias{vikhlinin2006b}.
\begin{equation}
    T(r) = T_0 \frac{(r/r_t)^{-a}}{\left[ 1 + (r/r_t)^b \right]^{c/b}} \frac{(r/r_{\text{cool}})^{a_{\text{cool}}} + T_{\text{min}}/T_0}{(r/r_{\text{cool}})^{a_{\text{cool}}} +1}
    \label{eqn:v06}    
\end{equation}
and a simplified form (which approximated an average temperature profile):
\begin{equation}
    \frac{T(r)}{T_{\text{mg}}} = \frac{(x/0.045)^{1.9} + 0.45}{(x/0.045)^{1.9} + 1} \frac{1.35}{(1 + (x/0.6)^2)^{0.45}},
    \label{eqn:v06avg}    
\end{equation}
where in Equation~\ref{eqn:v06}, all other variables except $r$ are fitted parameters, and in Equation~\ref{eqn:v06avg}, $x = r/r_{500}$ and $T_{\text{mg}}$ is the gas-mass-weighted temperature. In the general form, we opt to restrict the parameters $a=0$ and $b=2$, as in \citet{ghirardini2019} (The fitted curves with $a$ and $b$ fixed appear indistinguishable from the fitted curves when we vary $a$ and $b$ in the fits.). Figure~\ref{fig:zw3146_temperature} shows our temperature profiles along with these fitted parameterizations; the parameters themselves are listed in Table~\ref{tbl:temp_params}.

\begin{deluxetable*}{ccccccc}
\tabletypesize{\scriptsize}
\tablecolumns{10}
\tablewidth{0pt}
\tablecaption{Temperature Parameters \label{tbl:temp_params}}
\tablehead{ 
  \colhead{Model} & \colhead{$T_0$} & \colhead{$r_t$} & \colhead{$c$} & \colhead{$T_{\text{min}} / T_{0}$} & \colhead{$r_{\text{cool}}$} & \colhead{$a_{\text{cool}}$} \\
   & \colhead{(keV)} & \colhead{(Mpc)} & & & \colhead{(kpc)} & 
}
\startdata
    NP & 11.3 & 4.19 & 10 & 0 & 109 & 1.19 \\
    gNFW & 9.21 & 1.88 & 3.02 & 0.2 & 107 & 2.00 \\
    \hline \\
    priors & U(0,20) & U(0,5) & U(0,10) & U(0,1.0) & U(0,400) & U(0,10) 
\enddata
\tablecomments{Fitted parameters for temperature profiles to Equation~\ref{eqn:v06}, fixing $a=0$ and $b=2$. We apply limits to the fitted parameters; these are
expressed in the table as adding uniform priors between the lower, $l$, and upper, $u$, bounds: U($l$,$u$).}
\end{deluxetable*}

We note that Equation~\ref{eqn:v06} has more free parameters than are necessary to fit our temperature profiles.
The (approximate) average temperature profile (Equation~\ref{eqn:v06avg}) does not appear to be as good of a fit. In \citetalias{vikhlinin2006b}, they find one cluster (Abell 2390) which has a similar shape as ours; Abell 2390 was noted as having an unusual temperature profile
in their sample due to the central AGN in that cluster.

\subsubsection{Entropy}

Entropy is potentially equally important to pressure or temperature for studying the evolution of clusters. Additionally, as convective stability is achieved when $dK/dr \geq 0$, it determines the structure of the ICM \citep[e.g.][]{voit2005}.

We adopt the entropy parameter $K_e$ (hereafter ``entropy'') as defined in \citet{voit2005}, where:
\begin{equation}
    K_e = k_B T n_e^{-2/3} = P_e n_e^{-5/3}.
    \label{eqn:entropy}
\end{equation}
From \citet{voit2005} and references therein, we expect this proxy for entropy to follow as a function of radius a power law $K_e(r) \propto r^{1.1}$, with some deviation at small radii ($r \lesssim 0.1 r_{200}$), where the entropy profiles flatten towards the core. More recent works, \citep[e.g.][]{cavagnolo2009,pratt2010,ghirardini2019}, continue to find overall agreement with the so-called Voit profile, especially over sample averages. However, some differences, are found. \citet{cavagnolo2009} found a slightly steeper power-law slope ($\alpha = 1.21 \pm 0.39$) across the entire sample, which is still consistent with the Voit profile. Conversely, \citet{ghirardini2019} found a slope of $\alpha = 0.84 \pm 0.04$ at large radii. 

\begin{figure}
  \begin{center}
     \includegraphics[width=0.45\textwidth]{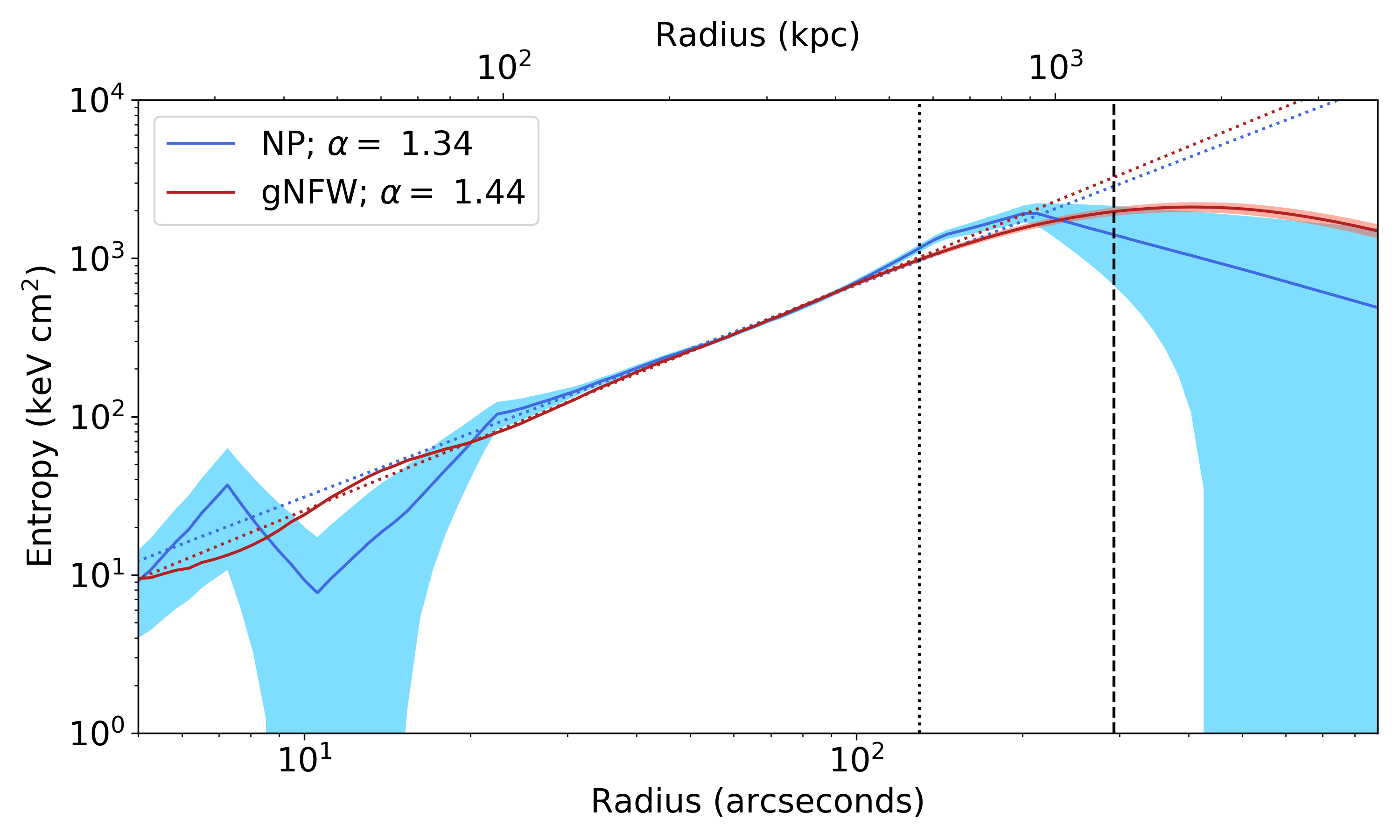}
  \end{center}
  \caption{Entropy profile from MUSTANG-2-derived pressure and \emph{XMM}-derived electron density, calculated as in Equation~\ref{eqn:entropy}. The error bars are solely statistical. The vertical lines are as in Figure~\ref{fig:zw3146_pressure_profiles}.}
  \label{fig:zw3146_entropy}
\end{figure}

Our entropy profiles, shown in Figure~\ref{fig:zw3146_entropy}, exhibit some noise in the central region in the NP fit, and show a turnover at large radii. We fit the power-law slope over the range $100 < r (\text{kpc}) < 600$ and found $\alpha_{\text{NP}} = 1.34 \pm 0.06$ and $\alpha_{\text{gNFW}} = 1.44 \pm 4e-5$. Note that by restricting the shape of the pressure profile, the gNFW parametrization reduces our uncertainty on the entropy slope.
For completeness, if we use all but the last bin in the NP model, (i.e. $r < 1000$ kpc), then the fitted slope is $\alpha = 1.43 \pm 0.04$.

\subsubsection{Gas fraction}

 We expect the matter content of galaxy clusters to be approximately those of the universe. That is, we expect $f_{gas}$ to be close to $\Omega_b / \Omega_m$, and measuring this can help us understand how matter is accreted and processed. Measuring $f_{gas}$ is notoriously difficult, certainly in the intergalactic medium (IGM). However, given the robust constraints on $f_{\text{gas}}$, especially from simulations, $f_{\text{gas}}$ can rather be used to gauge the non-thermal pressure support, when it is calculated with respect to the mass profile as derived from hydrostatic equilibrium. 

\begin{figure}
  \begin{center}
     \includegraphics[width=0.45\textwidth]{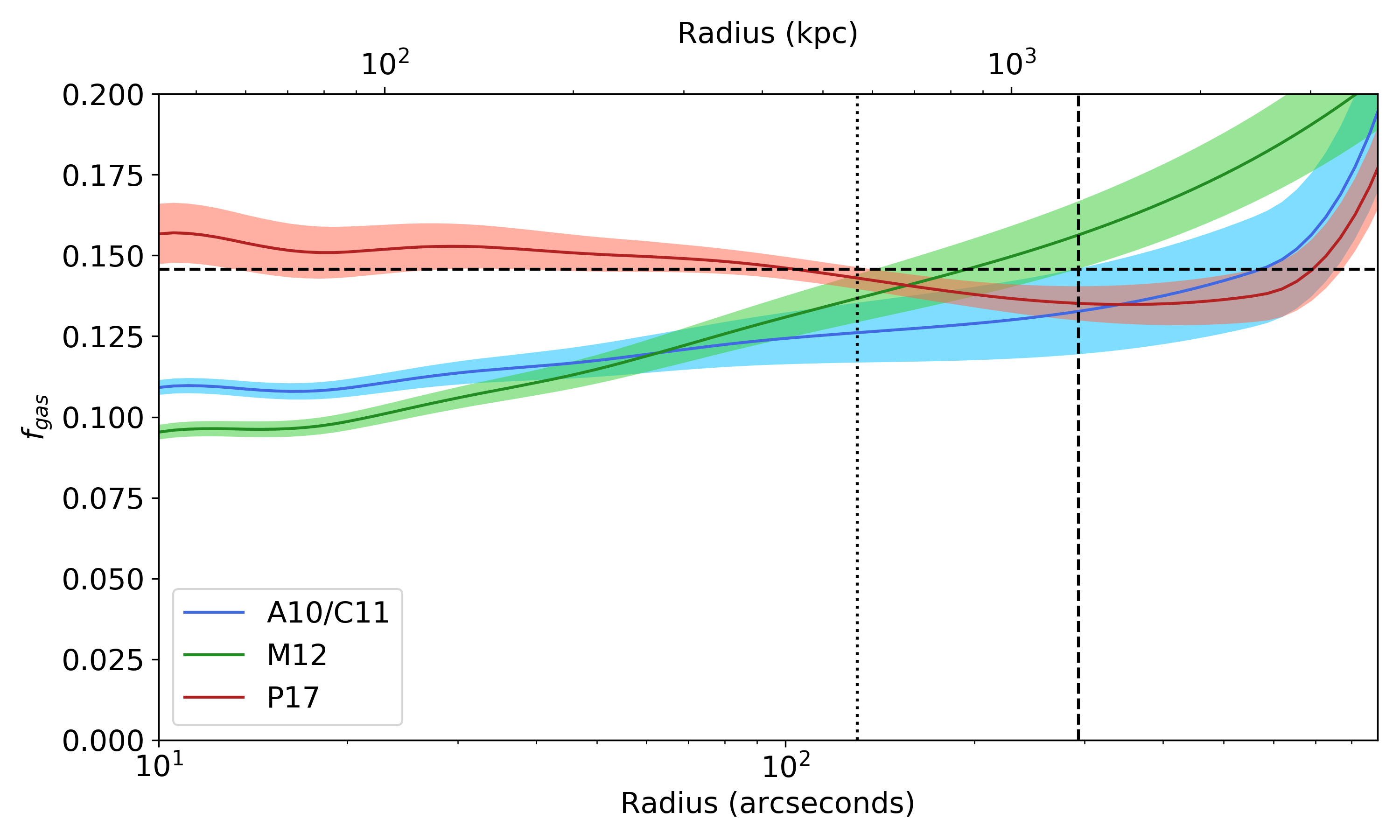}
     \includegraphics[width=0.45\textwidth]{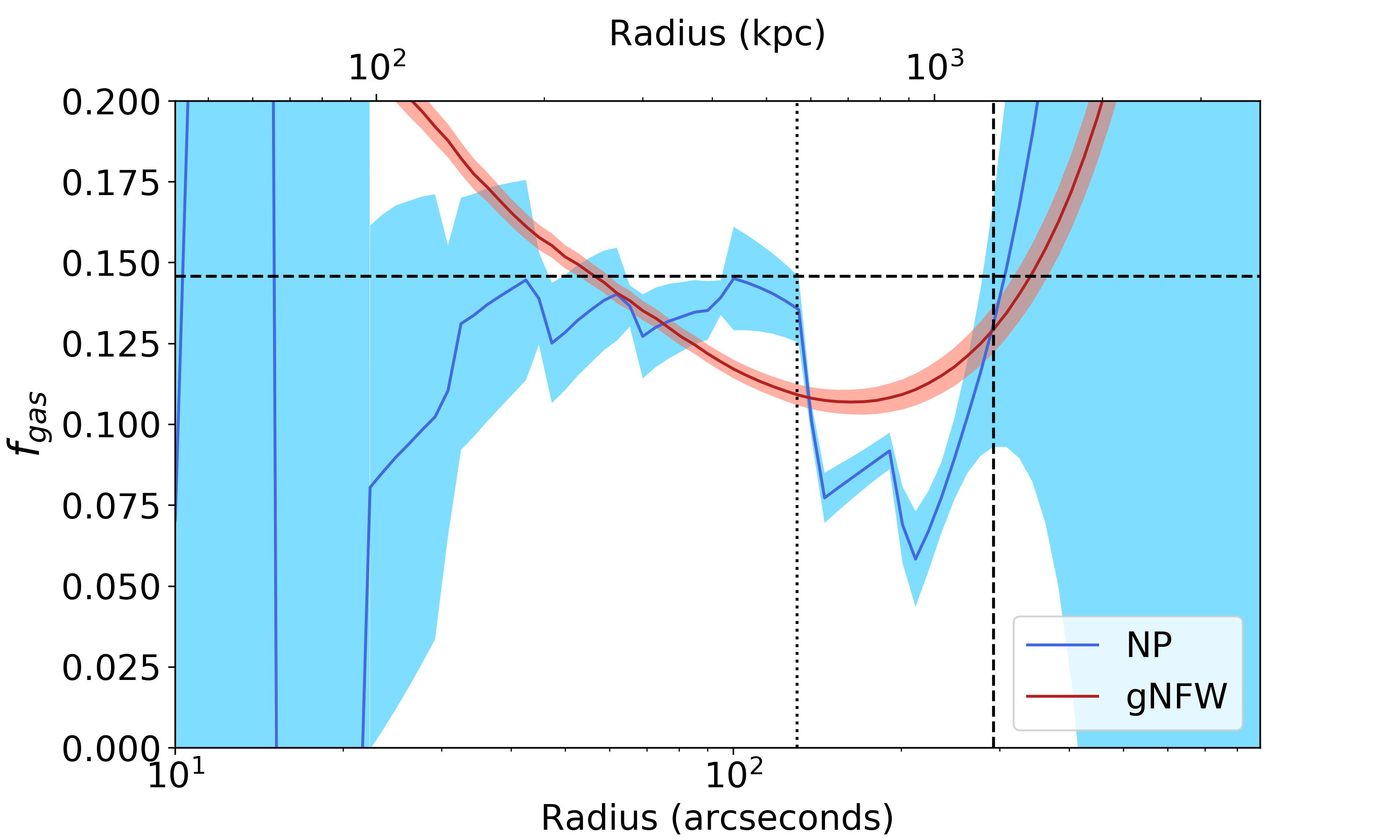}
  \end{center}
  \caption{Upper panel: gas fractions from $Y$-$M$ relations with the NP pressure profile model assuming an NFW mass profile and a constant mean molecular weight for the total mass. Bottom panel: $f_{\text{gas}}$ with respect to hydrostatic masses of the NP and gNFW models for the total mass.
  In both panels, the vertical lines are as in Figure~\ref{fig:zw3146_pressure_profiles}. The horizontal line is the universal cosmological baryon fraction, $\Omega_b / \Omega_m$.}
  \label{fig:zw3146_gas_fractions}
\end{figure}

In particular, we expect that $f_{gas} < \Omega_b/ \Omega_m$, largely due to formation processes \citep[e.g.][]{laroque2006} and a small amount of baryons will also be locked up in stars.
Figure~\ref{fig:zw3146_gas_fractions} shows the gas fractions as calculated with respect to total masses from $Y$-$M$ relations (upper panel) and from hydrostatic equilibrium (lower panel). The total mass profiles for the $Y$-$M$ relations fit a NFW mass profile to the respective mass pair of $M_{2500}$ and $M_{500}$ in the NP pressure profile model. Thus, the blue curve (A10/C11) in the top panel uses an NFW mass profile fitted to $M_{500} = 8.16^{+0.44}_{-0.54} \times 10^{14} \rm M_\odot$ and $M_{2500} = 3.68^{+0.41}_{-0.33} \times 10^{14} \rm M_\odot$. 

\section{Discussion}
\label{sec:discussion}
    
\subsection{SZ substructure}

Simulations consistently have shown that pressure, which underlies the Compton $y$, equilibrates more quickly than other thermodynamic parameters \citep[e.g.][]{motl2005, nagai2007}.  The typical timescale for pressure equilibration is characterized by the sound crossing time, which for most massive clusters is $\sim 1$~Gyr \citep[e.g.][]{sarazin1988, Sarazin2003}.  With no clear evidence of a recent, strong merger, it is not surprising that the residual signal (0.5\%) is less than a few percent of $Y_{cyl}(R_{500})$ since by most metrics this is a relaxed, cool core cluster. 

By way of comparison, we highlight the case of the well-known, strongly-sloshing X-ray luminous cluster RX~J1347.5-1145 \citep{komatsu2001,kitayama2004,kitayama2016,Ueda2017}. 
RX~J1347.5-1145 is a dramatic and likely late-stage merger that is arguably more disturbed than Zw3146 and has 2 clear lensing peaks traced by 2 equally-bright `brightest cluster galaxies' \citep[e.g.][]{Ueda2017}.
The excess SZ residual reported in \citet{korngut2011} and \citet{plagge2013}
and attributed to the SE enhancement is only $\approx 9-10\%$ of the total SZ signal within $R_{500}$. 
However, this estimate comes about when fixing the centroid to the X-ray peak and assuming spherical symmetry, rather than determining the centroid and geometry from the SZ data.
As shown in \citet{dimascolo2019}, an ellipsoidal pressure profile model with a floating centroid fits RX~J1347.5-1145 with no significant residuals across a broad range of scales and observations, including: the 12-meter ALMA in compact configuration ($\approx$5\arcsec\ resolution), the 7-meter Atacama Compact Array (ACA; $\approx$15\arcsec\ resolution), Bolocam ($\approx$1\arcmin\ resolution), and {\it Planck} ($\approx$10\arcmin\ resolution). It is therefore unsurprising Zw3146 can also be described well by a continuous pressure distribution, though we note again that in the case of Zw3146, the centroid choice does not strongly affect the inferred pressure profile (see Section \ref{sec:pres_profs}).  
 
We revisit briefly the ellipticity found in Zwicky 3146. \citet[][Section 3.4]{kravtsov2012} remark that isopotential surfaces in equilibrium are more spherical than the underlying mass distribution. Moreover, as baryons flow to the center, the underlying dark matter distribution is expected to be more spherical with decreasing cluster-centric radius \citep[][Section 3.5.3 and references therein]{kravtsov2012}. In \citet{romero2017}, the tendency for a more spherical core was also noted. Thus, the increase in ellipticity toward the center of Zwicky 3146 appears unusual and could be related to the sloshing in the core.
A more detailed study of the sloshing scenario in this cluster is in preparation.

\subsection{Recovery of pressure profile}
\label{sec:pp_recovery}

    Several previous ground-based single dish SZ studies have
    concluded
    that they cannot constrain the pressure profile of a cluster beyond the instrument's radial FOV (Section~\ref{sec:pres_profs}) or, similarly, they have concluded
    that an attempt to constrain the profile beyond the radial FOV may be biased \citep[e.g.][]{sayers2016}. Thus, achieving constraints on the pressure profile to better than 15\% out to 240\arcsec, or twice our radial FOV, is a marked improvement. 
    
    Our constraints diminish rapidly beyond 240\arcsec\ for multiple reasons: (1) our coverage drops off rapidly beyond this scale (due to the scanning radius used in mapping; see Figure \ref{fig:m2_scan_pattern}), (2) the cluster SZ signal is weaker, and (3) our surface brightness profile binning will be larger, and thus more susceptible to degeneracy with large-scale ($\gtrsim$ FOV) modes from the atmosphere. However, we note that, so long as the low-$k$ modes can be sufficiently sampled, Minkasi does not inherently limit the scales at which a model can be constrained.

    We find that the A10 profile (with $\alpha$, $\beta$, and $\gamma$ fixed, as specified in Section~\ref{sec:pres_profs}) fitted to our data are in good agreement with our non-parametric profiles. Although this is only one cluster, this acts as further indication that the A10 profile is a good descriptor of pressure profiles of galaxy clusters, especially relaxed galaxy clusters. By extension, we find that, at least for relaxed clusters, it is reasonable to fix the shape of the pressure profile to the A10 profile fit for a cluster when the data quality or angular coverage do not allow for proper non-parametric constraints (see also Appendix~\ref{sec:appendix_midas_minkasi}). In the case of Zwicky 3146, the deep MUSTANG-2 observations were critical to achieving the constraints on the non-parametric pressure profile that we found. 
    
    We have shown that it is possible to constrain the pressure profile non-parametrically beyond the FOV in SZ TODs. We found that our results are remarkably stable across a range of smoothing kernels when estimating the noise model. Furthermore, the agreement with the
    \emph{XMM-Newton}-derived pressure profile reinforces our confidence in the results. However, we note that some conflicts arise in products derived from our pressure profiles, such as the hydrostatic mass bias
    (Section~\ref{sec:dis_mass_est})
    and the entropy profiles (Section~\ref{sec:dis_thermo}).

\subsection{Mass Estimates}
\label{sec:dis_mass_est}

\subsubsection{Our mass estimators}
\label{sec:our_masses}

    We have estimated the mass from our pressure profiles via three estimators: (1) $Y$-$M$ scaling relations, (2) assuming HE in combination with X-ray determined electron densities, and (3) application of the virial theorem assuming uniform quantities $f_{\text{gas}}$ and $\mu_e$ and assuming that the matter profile follows a NFW profile. Where (2) and (3) produce mass profiles, (1) a single $Y$-$M$ relation is defined at a specific radius. Thus, we have chosen 6 $Y$-$M$ relations (3 at $R_{500}$ and 3 at $R_{2500}$) from \citetalias{arnaud2010,comis2011,marrone2012}, and \citetalias{planelles2017}. Across our mass estimates, we see a broad range; in particular, our minimum mass estimate is discrepant with our maximum estimate; similar results have been found in other works \citep[e.g.][]{hasselfield2013,Schrabback2018}, thus reaffirming that the choice of mass estimation method (e.g. assumed $Y$-$M$ relation) is important.

    Mass estimates from our gNFW (A10) pressure profile tend to be lower than our mass estimates, at both $M_{500}$ and $M_{2500}$ from our NP pressure profile. At $M_{500}$, the gNFW masses are in agreement with the NP masses as they are within 1$\sigma$, except for $M_{VT}$, where the difference is still within $2\sigma$.

    It is interesting that our hydrostatic masses at $R_{500}$ and $R_{2500}$ are at or above all respective $Y$-$M$ mass estimates. With the exception of \citetalias{arnaud2010} and \citetalias{comis2011}, which tie their masses to hydrostatic masses, the other $Y$-$M$ relations can trace their mass estimates back to either weak lensing \citepalias{marrone2012} or numerical simulations \citepalias{planelles2017}. 
    Mass estimates assuming thermal hydrostatic equilibrium are expected and generally found to be biased low because they do not account for non-thermal pressure support \citep[e.g.][]{ettori2011}. The hydrostatic mass bias is given by:
    \begin{equation}
        b = \frac{M_{\text{tot}} - M_{\text{HE}}}{M_{\text{tot}}},
    \end{equation}
    with $M_{\text{tot}}$ being the (true) total mass.
    The bias is typically found to be between 0.1 and 0.3 \citep[e.g.][and references therein]{hurier2018}. While \citet{ruppin2019} find that a few (individual) clusters have negative $b$; these are disturbed clusters. Rather, as in Figure 6 of \citet{ruppin2019}, all of the relaxed clusters have a positive $b$. 

    Of the $Y$-$M$ mass estimates at $R_{500}$, the P17 mass estimates (tied to simulation masses) appear consistent with the A10 (hydrostatic) mass estimates; however, M12, which comes from weak lensing, falls short of the other two  $Y$-$M$ mass estimates, as well as
    our HE and VT estimates. Similarly, at $R_{2500}$, the mass estimate from the M12 relation lies below the mass estimate from the other two $Y$-$M$ relations. The C11 mass is consistently above the P17 statistical error, but within the $Y$-$M$ systematic errors.
    
    Our mass estimates from the virial theorem, $M_{\text{VT}}$, have been calculated with the same assumptions as in \citet{mroczkowski2011}, and restricting the fitted region to radii outside the central 100 kpc owing to poorer constraints on the NP 
    pressure profiles. As indicated in Section~\ref{sec:results}, the outer bins (certainly in the MIDAS pipeline) may also adversely affect the fitted mass profile. Of the other assumptions made, the assumption of a fixed gas fraction may dominate the systematic error with this approach. 
    
\subsubsection{Comparison of our masses relative to previous values}
\label{sec:mass_compare}

\begin{figure}
        \begin{center}
        \includegraphics[width=0.45\textwidth]{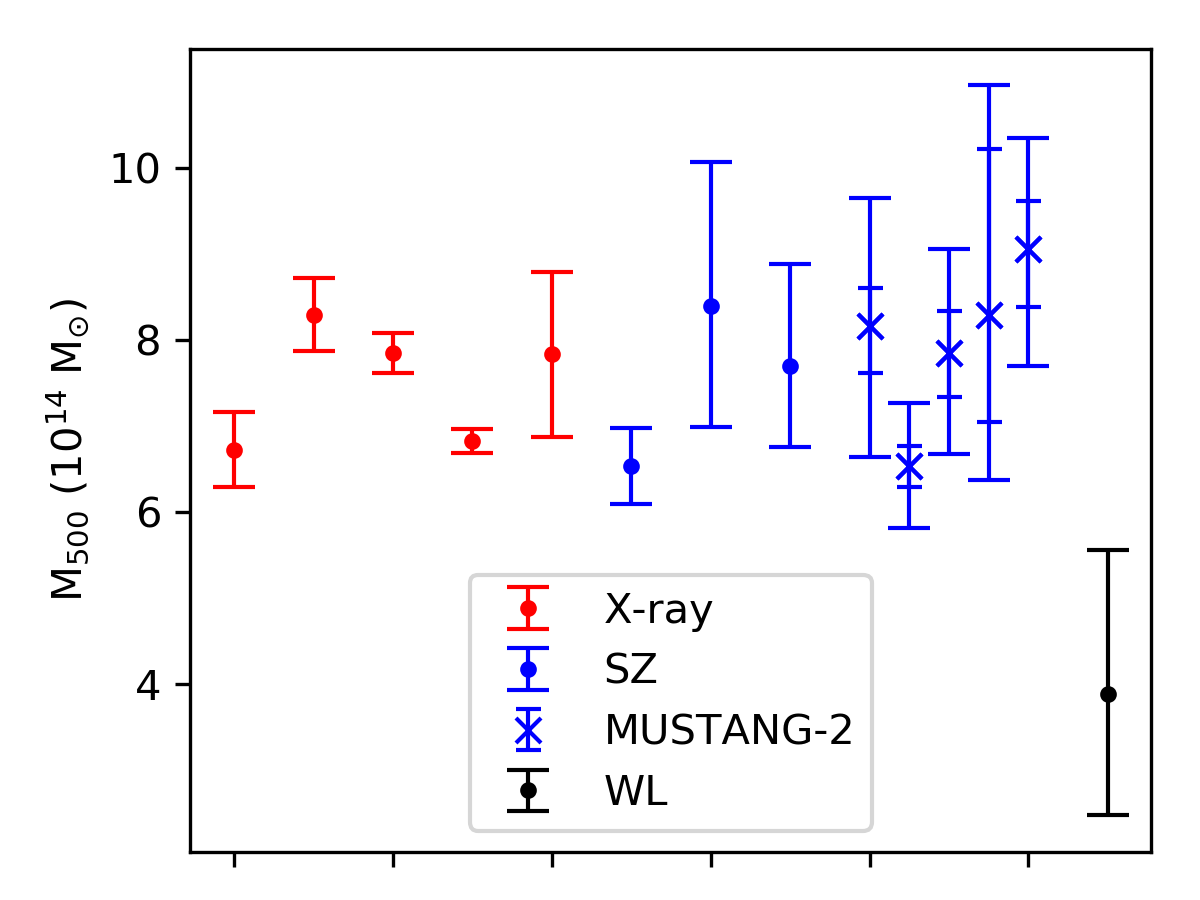}
        \end{center}
        \caption{Estimates of $M_{500}$ with uncertainties less than 20\% from Table~\ref{tbl:zwicky_masses} and color-coded based on the observation type. WL stands for weak lensing. Here, extrapolated $M_{500}$ values have their corresponding uncertainties extrapolated, if available. Some uncertainties from the literature include systematic errors while others do not, thus some caution is warranted when interpreting this plot.
        The MUSTANG-2 points show the statistical uncertainties (partial error bars) and statistical plus systematic uncertainties, added linearly (full error bars).
        }
        \label{fig:zw3146_lit_masses}
    \end{figure}

    As before, the Minkasi pipeline pressure profiles recover well the pressure profile beyond MUSTANG-2's FOV, and moreover, they are in good agreement with the pressure profile from \emph{XMM}.  
    However, the various methods of mass estimation have non-trivial scatter in an era when accuracy to better than 10\% \citep[e.g.][]{applegate2014,bocquet2019,miyatake2019} is the goal.
    Somewhat surprisingly, the $Y$-$M$ relation based on weak lensing masses (M12) appears the most discrepant among the other mass estimators at $R_{500}$. 
    One of the most recent mass estimate in the literature comes from weak lensing \citep{okabe2016}, and it too underestimates the mass relative to other mass estimates in the literature and those found in this work. The weak lensing mass estimate in \citet{kausch2007} also appears low relative to other mass estimates; only those derived from observations with the Nordic Optic Telescope \citep{dahle2006,pederson2007,sereno2015a} are around or above the median mass ($M_{500}$) reported. We present only mass estimates with uncertainties less than 20\% in Figure~\ref{fig:zw3146_lit_masses}, where we see broad agreement between the ICM-based masses. 
    Of the ICM-based mass estimates, we find the minimum and maximum come from MUSTANG-2. This highlights the importance of mass estimation method. We find that all ICM-based masses appear to scatter about $8 \times 10^{14}$ M$_\odot$. Of the recently derived masses from SZ, ACT and MUSTANG-2, in particular ($Y$-$M$)$_{1}$ in Table~\ref{tbl:m2_mass_Minkasi}, arrive at similar mass estimates, while Planck recovers a slightly lower mass than ACT and MUSTANG-2. These three mass estimates have used the same underlying $Y$-$M$ relation, and the three instruments have varying levels of detection significance. ACT detects Zwicky 3146 at $14\sigma$, Planck at $8.4\sigma$, and MUSTANG-2 at $104\sigma$. We note that our extrapolated value of $M_{500}$ from OVRO/BIMA is also in agreement with ACT, Planck, and MUSTANG-2.

    Given the agreement among ICM-based mass estimates and the expectation that ICM biases are minimized for relaxed clusters \citep[e.g.][]{biffi16}, the discrepancy between the ICM-based masses for Zwicky 3146 and the sole well-constrained (uncertainties $<20$\%) weak lensing mass estimate \citep{okabe2016} indicates a need for an improved understanding of our mass estimation methods and systematic uncertainties. In particular, the weak lensing mass estimate, being less than the ICM-based mass estimates, presents a challenging interpretation.

\subsection{Additional thermodynamic profiles}
\label{sec:dis_thermo}

Beyond mass estimates, we use the SZ-derived pressure profiles in combination with the X-ray-derived electron density to calculate temperature profiles, entropy profiles, and gas (mass) fraction profiles. These profiles are of intrinsic interest and also serve to further check consistency.

From the temperature profile, we find that temperature begins to decrease around $R_{2500}$, which is slightly larger than what was found for the average temperature profile in \citetalias{vikhlinin2006b}. The temperature does appear to drop quite low beyond $R_{500}$ for both our NP and gNFW models (Figure~\ref{fig:zw3146_temperature}). 

The drop in temperature at the outermost radii has a more pronounced consequence in the entropy profile. The entropy profile is expected to follow a power law outside of the core; simulations predict it will only turn over beyond $R_{200}$, while a handful of observations, largely of merging systems, show a turnover at and perhaps just interior to $R_{200}$ \citep[e.g.][]{walker2019}. Thus, the turnover just before $R_{500}$ appears suspect, especially as it only concerns one pressure profile bin. In Appendix~\ref{sec:appendix_adjustments} we investigate the impact on mass estimation if the entropy trend continued along its power law ($\alpha = 1.34$ for the NP model) in the event that this outermost bin is erroneously biased low in our data processing.

\section{Conclusions}
\label{sec:conclusions}
    We have taken deep observations of Zwicky 3146 with MUSTANG-2. These observations have allowed us to produce non-parametric constraints on the pressure profile in the radial range $5^{\prime\prime} < r < 300^{\prime\prime}$ through a newly developed processing pipeline, dubbed Minkasi. The pressure profile recovered is in excellent agreement with the pressure profile derived from \emph{XMM}.

Our $M_{500}$ estimates are fairly self-consistent. With our non-parametric model, excluding the $Y$-$M$ estimate from M12, our estimates of $M_{500}$ differ by less than 16\%. Despite this apparent consistency, we call attention to the fact that that the hydrostatic mass is slightly above the total mass as calculated from the $Y$-$M$ relations, which implies a negative hydrostatic mass bias. A positive mass bias (predominantly due to non-thermal pressure support) between 0.1 and 0.3 is generally expected from relaxed to merging systems (Section \ref{sec:discussion}).

Analysis of other thermodynamic quantities, such as the temperature profile and entropy profile, along with the hydrostatic mass profile suggests that the pressure close to and beyond $R_{500}$ may be underestimated. Within the NP model, we adjust the outermost pressure bin such that the entropy profile continues on a straight power-law. With this adjustment, the hydrostatic mass biases become more negative. Thus, while the adjustment presents a more coherent picture with respect to hydrostatic mass profile and entropy profile, it conversely indicates a potential problem in all three $Y$-$M$ relations, the resolution to which will require a systematic study of a large statistically-representative sample spanning a range of masses and evolutionary states.

We take our fiducial mass to be $8.16^{+0.44}_{-0.54}$ ($\pm 5.5$\% stat) ${}^{+0.46}_{-0.43}$ ($\pm 5.5$\% sys, $Y$-$M$) ${}^{+0.59}_{-0.55}$ ($\pm 7.0$\% sys, cal.) In light of the above discussion, we note that there is an additional systematic error due the issues discussed in the previous two paragraphs, but which we are currently unable to quantify. We see that our value is well within the range of reported values in the literature. It is in closer agreement to other studies based off the ICM than to weak lensing studies, which have tended to favor lower masses for Zwicky 3146. Including all the systematic uncertainties, this corresponds to only a $2\sigma$ discrepancy with the latest weak lensing measurement \citep{okabe2016}. This tension is but one of many that can be found among mass estimates in the literature for Zwicky 3146. Together these tensions demonstrate that systematic mass uncertainties are still non-negligible even in the so-called era of precision cluster cosmology.

Part of exploring the tensions in mass estimates will be a more thorough analysis of the physics within Zwicky 3146. In particular, we have begun a more detailed investigation into sloshing and pressure fluctuations within Zwicky 3146. These investigations aim to directly constrain the non-thermal pressure support. Beyond these additional analyses of the ICM physics, we find that Zwicky 3146 is ripe for further weak lensing analyses, which may help address offsets in $Y$-$M$ relations such as those from \citetalias{marrone2012} and \citetalias{arnaud2010}.

\section{Acknowledgements}

This work is supported by the NSF award number 1615604 and by the Mt.\ Cuba Astronomical Foundation. 
Sara Stanchfield is supported by NASA NSTRF NNX14AN63H.
Craig Sarazin is supported in part by {\it Chandra} grants GO7-18122X/GO8-19106X and {\it XMM-Newton} grants NNX17AC69G/80NSSC18K0488.
Massimo Gaspari is supported by the {\it Lyman Spitzer Jr.} Fellowship (Princeton University) and by NASA {\it Chandra} grants GO7-18121X/GO8-19104X/GO9-20114X.
This paper makes use of the following ALMA data: ADS/JAO.ALMA\#2011.0.00001.CAL. 
ALMA is a partnership of ESO (representing its member states), NSF (USA) and NINS (Japan), together with NRC (Canada), MOST and ASIAA (Taiwan), and KASI (Republic of Korea), in cooperation with the Republic of Chile. The Joint ALMA Observatory is operated by ESO, AUI/NRAO and NAOJ.
The National Radio Astronomy Observatory is a facility of the National Science Foundation operated under cooperative agreement by Associated Universities, Inc. GBT data was taken under the project ID AGBT18A\_175. Basic research in radio astronomy at the Naval Research Laboratory is
supported by 6.1 Base funding.
The authors would like to thank Stefano Andreon and Eric Murphy for valuable input provided.

\facilities{GBT, XMM}
\software{ds9,IDL,Astropy}

\bibliographystyle{aasjournal}
\bibliography{references}

\appendix
\section{appendix Self Consistency}
\label{sec:appendix_self_consistency}
\begin{figure}
  \begin{center}
     \includegraphics[width=0.45\textwidth]{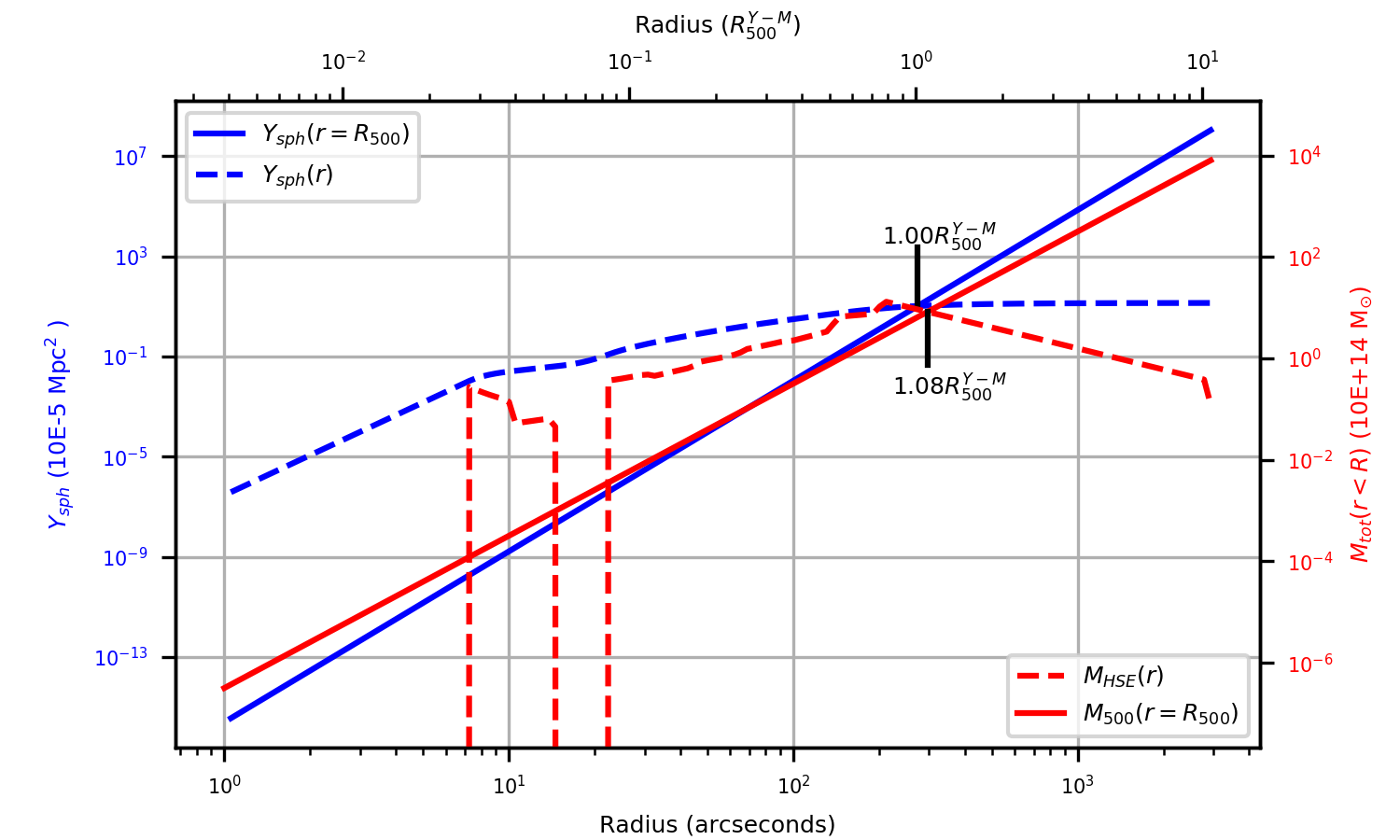}
  \end{center}
    \label{fig:zw3146_integrated_quantities}
  \caption{Hydrostatic mass is found at the point where $M_{tot}(<r)$ crosses $M_{500} = \rho_c 500 4 \pi r^3 / 3$. Similarly, the self-consistent $Y$ value is found where it crosses. Iteration is not necessary when comparing to a reference curve.}
\end{figure}

    Computing $M_{500}$ is equivalent to finding $R_{500}$ as $M_{500} = (4 \pi / 3) \times 500 \times \rho_c \times R_{500}^3$. Any self-consistent $M_{500}$ must then lie along the curve for all possible $R_{500}$ values (i.e. all radii); see Figure~\ref{fig:zw3146_integrated_quantities}. For mass estimates from hydrostatic equilibrium (dashed red curve) and the Virial theorem, where we have mass curves, we simply find where those mass curves cross a reference $M_{500}$ curve (solid red line).

By extension, a reference $Y_{sph}$ curve (line in log-log space) is created from the reference $M_{500}$ curve and the selected $Y$-$M$ relations, as expressed by:
\begin{equation}
    h(z)^{-2/3} Y_{sph} = 10^{A} \times M^{B},
    \label{eqn:YMrel}
\end{equation}
where $A$ is the logarithmic normalization and $B$ is the logarithmic slope. The relations used are those found in \citetalias{arnaud2010,comis2011,marrone2012,planelles2017}, and are tabulated in Table~\ref{tbl:YM_params}. In Figure~\ref{fig:zw3146_integrated_quantities}, a reference $Y_{sph}$ curve is the solid blue line, and the measured $Y_{sph}$ is given by the dashed blue curve.

We see that the hydrostatic mass curve is subject to spurious (negative) masses where the pressure increases. This is seen at small radii where the number of independent measurements is small and radio sources contribute to weaker constraints. Thus, these small radii are excluded when finding where the red curves (dashed and solid) intersect. The measured $Y_{sph}$ curve is an integrated quantity, and is comparatively well-behaved.

\begin{deluxetable}{ccc}
\tabletypesize{\scriptsize}
\tablecolumns{3}
\tablewidth{0pt}
\tablecaption{Temperature Parameters \label{tbl:YM_params}}
\tablehead{ 
  \colhead{Relation} & \colhead{A} & \colhead{B}
}
\startdata
    ($Y$-$M$)$_{\textsc{A10}}$ ($R_{500}$) & 1.78  & -30.515 \\
    ($Y$-$M$)$_{\textsc{M12}}$ ($R_{500}$) & 2.273 & -28.735 \\
    ($Y$-$M$)$_{\textsc{P17}}$ ($R_{500}$) & 1.685 & -29.073 \\
    \hline
    ($Y$-$M$)$_{\textsc{C11}}$ ($R_{2500}$) & 1.637 & -28.13 \\
    ($Y$-$M$)$_{\textsc{M12}}$ ($R_{2500}$) & 1.818 & -30.669 \\
    ($Y$-$M$)$_{\textsc{P17}}$ ($R_{2500}$) & 1.755 & -29.683 
\enddata
\tablecomments{$Y_{sph}$-$M$ relations expressed in the form of Equation~\ref{eqn:YMrel}. The relations are taken from \citetalias{arnaud2010,comis2011,marrone2012,planelles2017}.}

\end{deluxetable}

\section{appendix Profile Adjustments}
\label{sec:appendix_adjustments}
    
While we recover the pressure profile beyond the FOV very well, the very last bin appears to be biased low, as evidenced by our entropy profile (Sections~\ref{sec:results} and \ref{sec:discussion}). We thus investigate how our results change if we assume that our entropy profile should continue as a power law through the last radial bin as had been used in the NP pressure profile.
To do this, we perform two separate adjustments: first, we modify just the outermost pressure bin, and second (using the original pressure profile), we modify the corresponding electron density bin when matched to the pressure profile binning.

\begin{figure}
  \begin{center}
     \includegraphics[width=0.45\textwidth]{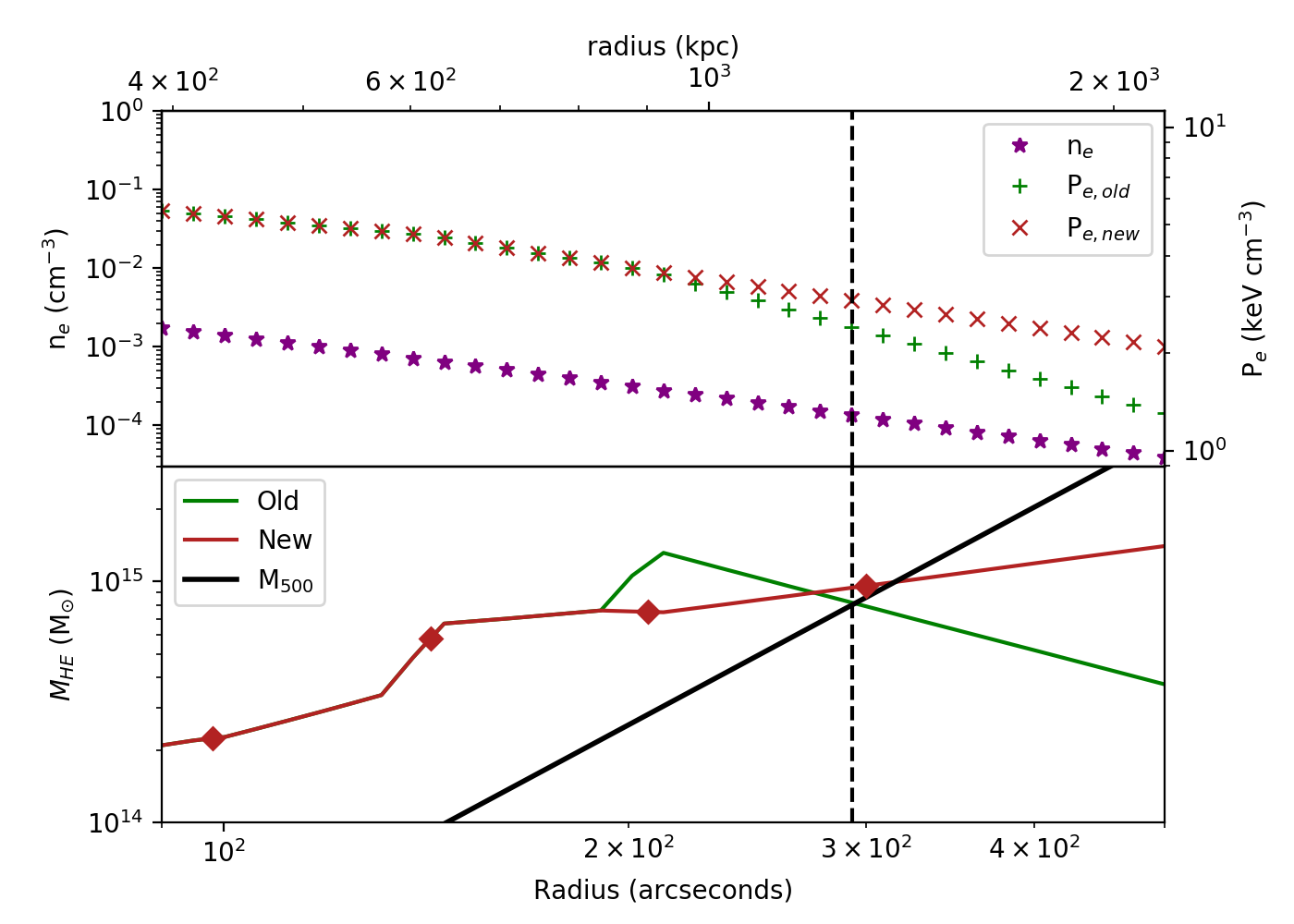}
  \end{center}
  \caption{Using the entropy profile (Figure~\ref{fig:zw3146_entropy}) as a guide to adjust the outermost pressure profile bin (and thus the outer slope), we obtain a new pressure profile (top panel, green X's). The bottom panel shows the old and new $M_{\text{HE}}$ profiles. Note that while our pressure profiles are fitted at 12 radii (bins), we have plotted with an interpolated binning.}
  \label{fig:zw3146_adjusted_Mhse}
\end{figure}

Figure~\ref{fig:zw3146_adjusted_Mhse} shows the adjusted pressure profile (with original electron density and original (old) pressure profile for reference) in the upper panel, and the subsequent hydrostatic mass profile (with old $M_{\text{HE}}$ profile in red and $M_{500} = 4 \pi \rho_c 500 \, r^3 / 3$ line in black for reference). The full set of resultant (adjusted) masses are reported in Table~\ref{tbl:m2_adj_mass}. When adjusting the outermost pressure bin, we find that the new pressure value is 2.3$\times$ higher than it was originally. This corresponds to an upward shift of $2.1\sigma$. 

\begin{deluxetable*}{l | c c c c c c}
\tabletypesize{\scriptsize}
\tablecolumns{7}
\tablewidth{0pt}
\tablecaption{Adjusted $M_{500}$ Mass Estimates \label{tbl:m2_adj_mass}}
\tablehead{\colhead{Adjustment} & \colhead{Quantity} & \colhead{($Y$-$M$)$_{\text{A10}}$} & \colhead{($Y$-$M$)$_{\text{M12}}$} & \colhead{($Y$-$M$)$_{\text{P17}}$} & \colhead{$M_{\textsc{HE}}$} & \colhead{$M_{\textsc{VT}}$} \\
\colhead{} & \colhead{} & \colhead{($10^{14} M_{\odot}$)} & \colhead{($10^{14} M_{\odot}$)} & \colhead{($10^{14} M_{\odot}$)} & \colhead{($10^{14} M_{\odot}$)} & \colhead{($10^{14} M_{\odot}$)}}
\startdata
    \hline 
        & & & & & & \\
    \multirow{3}{*}{None} & $M_{500}$ ($10^{14} M_{\odot}$) & $8.16^{+0.44}_{-0.54} {}^{+0.46}_{-0.43} {}^{+0.59}_{-0.55}$ & $6.53^{+0.24}_{-0.24} {}^{+0.14}_{-0.14} {}^{+0.36}_{-0.34}$
                            & $7.85^{+0.49}_{-0.52} {}^{+0.10}_{-0.10} {}^{+0.61}_{-0.56}$ & $8.29^{+1.93}_{-1.24} {}^{+0.74}_{-0.68}$ & $9.05^{+0.56}_{-0.67} {}^{+0.74}_{-0.68}$ \\
        & & & & & & \\
        & $b_{\text{old}}$ & -0.01 & -0.27 & -0.05 & -- & -- \\
        \hline
    \multirow{2}{*}{New $P_e$} & $M_{500}$ ($10^{14} M_{\odot}$) & $8.52$ & $6.71$ & $8.38$ & $9.95$ & $9.61$ \\
        & $b_{\text{new}}$ & -0.17 & -0.48 & -0.19 & -- & -- \\
        \hline
    \multirow{2}{*}{New $n_e$} & $M_{500}$ ($10^{14} M_{\odot}$) & $8.16$ & $6.53$ & $7.85$ & $12.7$ & $9.61$ \\
        & $b_{\text{new}}$ & -0.56 & -0.94 & -0.61 & -- & -- \\
        \hline
\enddata
\tablecomments{Resultant masses when adjusting the pressure in the last bin of the NP model. As the "new" masses are not derived directly from the (true) data, the error bars are not reported. The hydrostatic mass bias, $b$ is calculated as $(M_{Y\text{-}M} - M_{\text{HE}})/M_{Y\text{-}M}$ for the respective $Y$-$M$ relation. We note that ($Y$-$M$)$_{\text{A10}}$ is derived from hydrostatic masses, and thus the respective value for $b$ should be close to zero.}
\end{deluxetable*}

Indeed, the $Y$-$M$ masses increase with adjustment in pressure. Yet, the hydrostatic masses increase more than the $Y$-$M$ masses, and thus the hydrostatic mass biases, $b$, are driven further negative (Table~\ref{tbl:m2_adj_mass})! The hydrostatic masses are driven exceptionally low when adjusting the electron density. Clumping, quantified by $C = \frac{\langle n_e^2 \rangle}{\langle n_e \rangle^2}$, is a potential systematic for X-ray observations and would tend to overestimate the electron density. However, given that a diminished electron density in the outer radii dramatically worsens the hydrostatic mass bias, this does not appear to be an explanation. 

Figure~\ref{fig:zw3146_adjusted_Mhse} also reveals that the ``old'' $M_{\text{HE}}$ profile shows a decreasing mass at the largest radii. In combination with the entropy profile, this is quite suggestive that our last pressure profile bin is biased low. This is not too surprising given the lesser coverage and weaker constraints on noise at these low-k modes in the MUSTANG-2 TODs.
However, as above, a naive resolution to these symptoms results in more negative hydrostatic mass biases. 
That is, by extension, there is an additional problem to be solved - potentially it lies in all of the three $Y$-$M$ relations. However, this is beyond the scope of this paper.

\section{appendix Comparing MIDAS and Minkasi}
\label{sec:appendix_midas_minkasi}
\begin{figure}
    \begin{center}
    \includegraphics[width=0.45\textwidth]{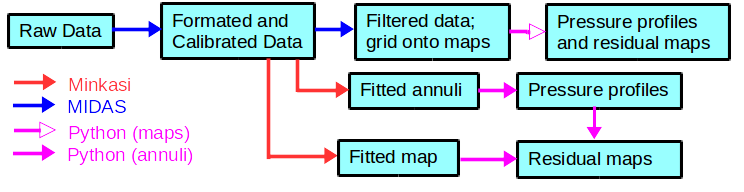}
    \end{center}
    \caption{Flowchart for how various data products are produced.}
    \label{fig:idl_mink_flowchart}
  \end{figure}

     While this work has focused on the results from Minkasi data processing, we have performed much of the same analysis through our other approach, MIDAS (MUSTANG IDL Data Analysis System). MIDAS benefits from its legacy of use with MUSTANG; its performance is well understood across a range of observational strategies, and thus still commonly used. In light of this, a comparison of performance is prudent, and follows below. For completeleness, Figure~\ref{fig:idl_mink_flowchart} is a flowchart of the data reduction branches.
  
  \subsubsection{Data processing with MIDAS}
  
  For each scan:
  
  (1) A pixel mask is defined based on responsivity of detectors from instrument setup at the beginning of the run; unresponsive detectors are masked out.
  
  (2) For each scan, the TODs are read in (2ms integrations) and averaged to larger time bins (20 ms integrations). Often, we apply a high- and low-pass filter to our TODs. The high-pass filter is chosen to filter out scales larger than the FOV (given our scanning speed). Similarly, the low-pass filter is chosen to filter out scales much smaller than our resolution. For this work, we adopted a high-pass filter at 0.08 Hz and a low-pass filter at 41 Hz.
  
  (3) Gain and opacity corrections are applied to our data.
  
  (4) Noise templates are constructed and fit to the data. The simplest form is to have one common mode across all detectors (typically the median). In addition to a common mode template, a high order ($\sim20$) order polynomial is simultaneously fit. The fits are done per detector, and subsequently the fitted templates are subtracted. Alternatively the template may be the $N$ ($N$ is chosen by the user; usually between 2 and 5) principal components of TODs. 
  
  (5) The cleaned TODs are checked for glitches, where a small portion of the TOD (from just before to just after each glitch) is flagged. Detector weights are assigned based on the RMS of the corresponding TOD.
  
  (6) The cleaned TODs are passed to a gridding routine. A data map and a weight map are created. For this cluster, we use 2\arcsec\ pixels.
  
\begin{figure}
  \begin{center}
     \includegraphics[width=0.45\textwidth]{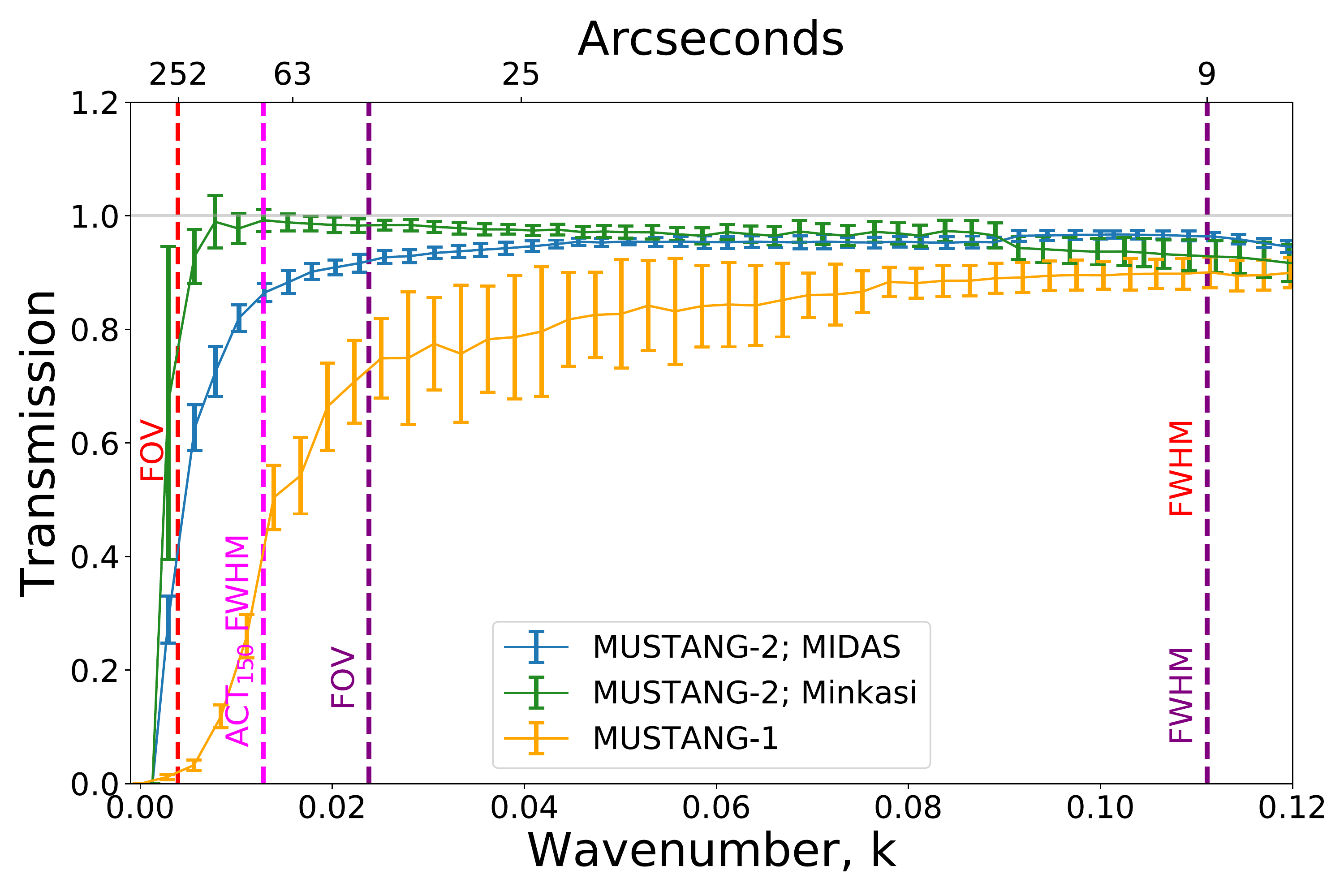}
    \includegraphics[width=0.45\textwidth]{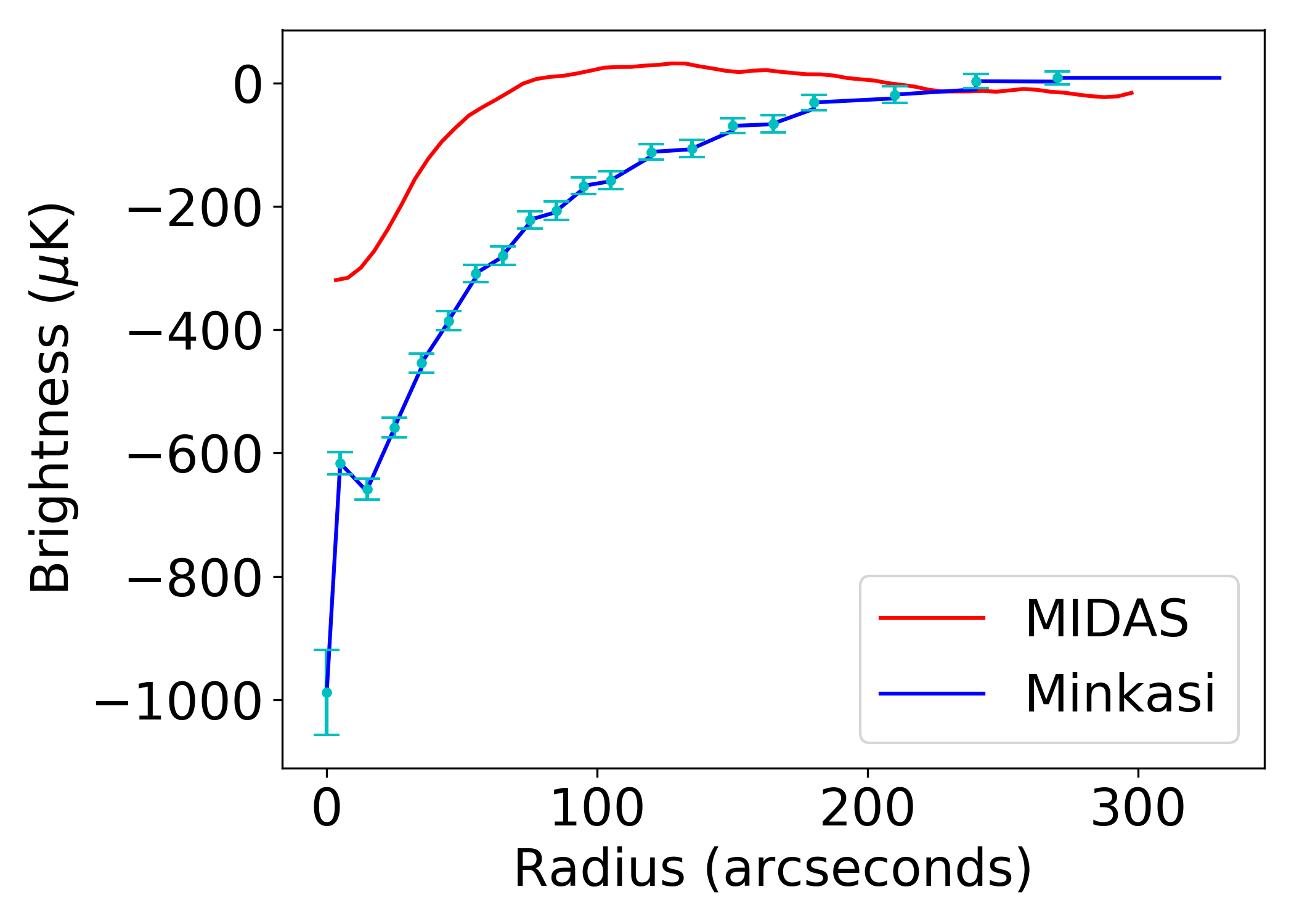}
  \end{center}
    \label{fig:m2_xfer}
  \caption{Top: the transfer function of MUSTANG-2 shows a marked improvement over that of MUSTANG-1. We also note the overlap with ACT (especially the band centered at 146 GHz, noted as 150 in this figure). Bottom: the surface brightness profile for Zwicky 3146 as determined by Minkasi (in concentric annuli, as plotted) and by MIDAS; the impact of the transfer function(s) is evident.}
\end{figure}

  \subsubsection{Comparing the performance of the two methods of data processing}
    
    Within MIDAS, for Zwicky 3146 we find that low pass filters between 0.06~Hz and 0.09~Hz sufficiently reduce the atmospheric signal, while still retaining much of the SZ signal. Our final processing uses a low pass at 0.08~Hz. Modulation of other processing parameters has comparatively minor effects on the resultant map(s).

    The recovered signal between MIDAS and Minkasi is illustrated via the transfer function (left panel) and surface brightness profile with point sources removed (right panel) in Figure~\ref{fig:m2_xfer}. While the bulk of the difference in the recovered surface brightness profiles is due to the difference in transfer functions, we also note that the Minkasi surface brightness profile is deconvolved (from the MUSTANG-2 beam), while the MIDAS surface brightness profile is not.
    Within MIDAS, the sources are modelled as point sources with the average beam for all observing nights, taken as a single spherically-symmetric Gaussian with FWHM of 10\arcsec\!.7.

\subsection{Pressure Profile Fitting and Results}

\begin{figure}
  \begin{center}
     \includegraphics[width=0.45\textwidth]{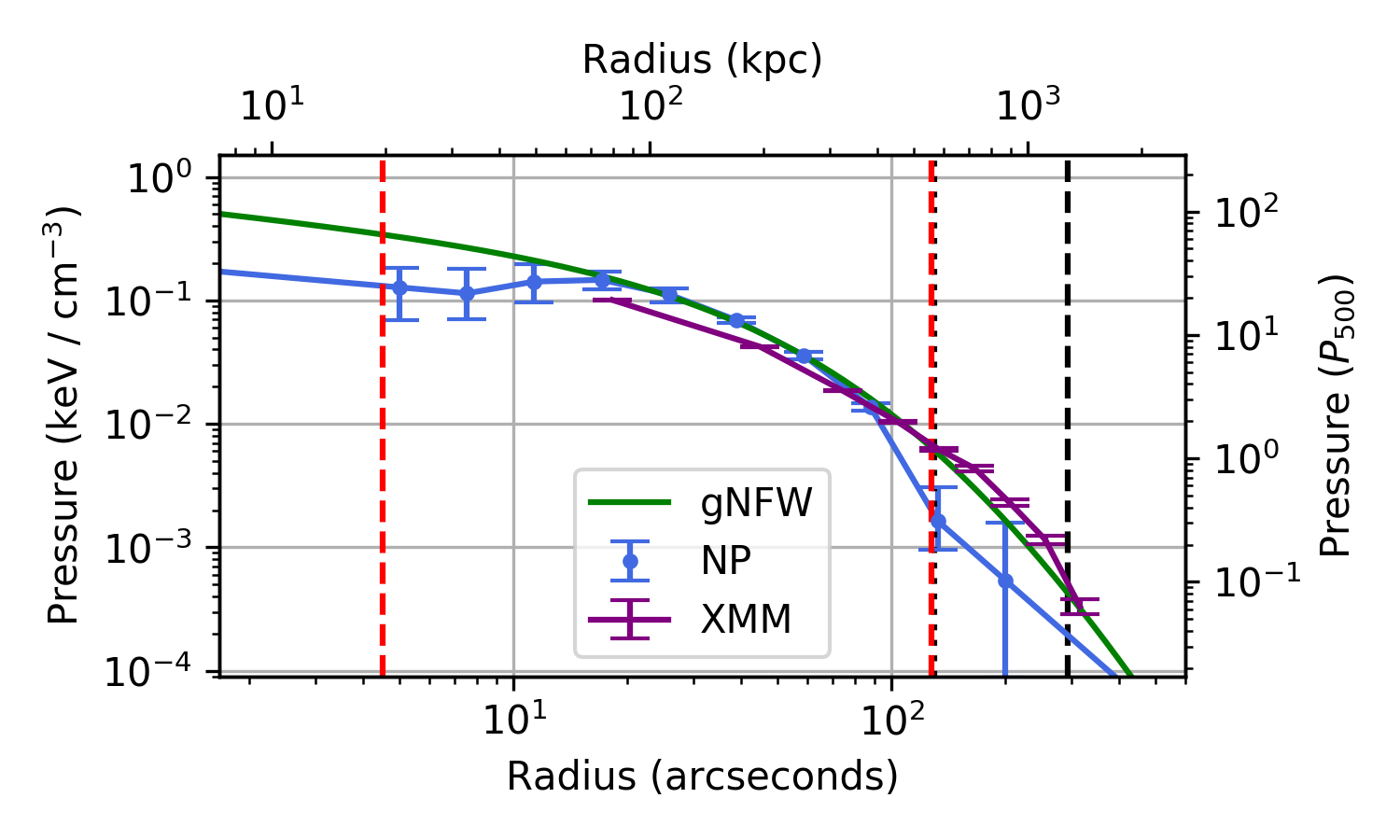}
  \end{center}
  \caption{The pressure profiles for Zwicky 3146 as recovered by MIDAS, as well as the pressure profile as determined from \emph{XMM-Newton}. 
  The vertical red dashed lines are the HWHM and radial FOV for MUSTANG-2; the vertical black dashed line is $R_{500}$ for $M_{500} = 8\times 10^{14} M_{\odot}$ and the vertical black dotted line is $R_{2500}$ for $M_{2500} = 3.5\times 10^{14} M_{\odot}$.}
  \label{fig:midas_pp}
\end{figure}


\begin{deluxetable*}{l | c c c c c c}
\tabletypesize{\scriptsize}
\tablecolumns{5}
\tablewidth{0pt}
\tablecaption{MUSTANG-2 Mass Estimates from MIDAS \label{tbl:m2_mass_MIDAS}}
\tablehead{\colhead{$M_{\Delta}$} & \colhead{Model} & \colhead{($Y$-$M$)$_1$} & \colhead{($Y$-$M$)$_2$} & \colhead{($Y$-$M$)$_3$} & \colhead{$M_{\textsc{HE}}$} & \colhead{$M_{\textsc{VT}}$} \\
 & \colhead{} & \colhead{($10^{14} M_{\odot}$)} & \colhead{($10^{14} M_{\odot}$)} & \colhead{($10^{14} M_{\odot}$)} & \colhead{($10^{14} M_{\odot}$)} & \colhead{($10^{14} M_{\odot}$)}}
\startdata
        & & & & & & \\
    \multirow{3}{*}{$M_{500}$}     & NP & $5.72^{+1.05}_{-1.03} {}^{+0.24}_{-0.24} {}^{+0.42}_{-0.39}$ & $5.02^{+0.54}_{-0.49} {}^{+0.20}_{-0.19} {}^{+0.27}_{-0.26}$
                             & $5.32^{+0.85}_{-0.91} {}^{+0.07}_{-0.07} {}^{+0.41}_{-0.38}$ & $0.87^{+0.05}_{-0.32} {}^{+2.97}_{-3.08}$ & $14.7^{+6.4}_{-4.9} {}^{+1.2}_{-1.1}$ \\
        & & & & & & \\
                             & gNFW & $7.38^{+0.15}_{-0.14} {}^{+0.39}_{-0.37} {}^{+0.54}_{-0.50}$ & $6.01^{+0.09}_{-0.09} {}^{+0.22}_{-0.21} {}^{+0.33}_{-0.31}$
                             & $7.06^{+0.15}_{-0.15} {}^{+0.09}_{-0.09} {}^{+0.55}_{-0.51}$ & $5.73^{+0.26}_{-0.28} {}^{+0.19}_{-0.17}$ & $7.52^{+0.14}_{-0.13} {}^{+0.56}_{-0.52}$ \\
        & & & & & & \\
    \multirow{3}{*}{$M_{2500}$}     & NP & $3.07^{+0.12}_{-0.10} {}^{+0.11}_{-0.11} {}^{+0.25}_{-0.23}$ & $2.73^{+0.07}_{-0.07} {}^{+0.15}_{-0.14} {}^{+0.19}_{-0.18}$
                             & $2.45^{+0.07}_{-0.06} {}^{+0.03}_{-0.03} {}^{+0.18}_{-0.17}$ & $2.86^{+0.16}_{-0.09} {}^{+7.39}_{-0.21}$ & $4.98^{+1.13}_{-1.32} {}^{+0.51}_{-0.46}$ \\
        & & & & & & \\
                             & gNFW & $3.69^{+0.06}_{-0.06} {}^{+0.15}_{-0.15} {}^{+0.30}_{-0.28}$ & $3.16^{+0.04}_{-0.04} {}^{+0.08}_{-0.08} {}^{+0.22}_{-0.21}$
                             & $2.82^{+0.04}_{-0.04} {}^{+0.03}_{-0.03} {}^{+0.21}_{-0.19}$ & $4.58^{+0.17}_{-0.18} {}^{+0.13}_{-0.10}$ & $3.63^{+0.05}_{-0.05} {}^{+0.31}_{-0.29}$ \\
        & & & & & & \\
\enddata
\tablecomments{Mass estimates from MUSTANG-2 (and electron density profiles from \emph{XMM} for $M_{\text{HE}}$. $M_{\text{VT}}$ is not the virial mass as is classically defined (with respect to $R_{vir}$), but rather the mass within $R_{\Delta}$ using the Virial theorem \citep{mroczkowski2011}. The error bars on the $Y$-$M$ mass are, in order, the statistical error, systematic error from the $Y$-$M$ relation itself, and the systematic error due to calibration uncertainty. The error bars on the other mass estimates are the statistical and systematic error due to calibration uncertainty. The $Y$-$M$ relations for $M_{500}$ for the subscripts 1, 2, and 3 are \citetalias{arnaud2010,marrone2012,planelles2017}, respectively; for $M_{2500}$ the relations come from \citetalias{comis2011,marrone2012,planelles2017}, respectively.}
\end{deluxetable*}

Pressure profile fitting to MIDAS maps is very similar to that in Minkasi. The same line-of-sight integration schemes are used. The major difference is that once we have calculated a Compton $y$ profile and converted it to brightness temperature, $T_\textsc{B}$, we must grid it onto our map, convolve by the MUSTANG-2 beam, and apply our transfer function (the blue curve in Figure~\ref{fig:m2_xfer}). We simultaneously the six point sources, which also have the transfer function applied to them.  Given our reduced transmission at large scales with MIDAS, we use 10 bins logarithmically spaced between 5\arcsec\ and 3\arcmin\ in our NP model.

 The non-parametric constraints in the MIDAS profile show a drop at our radial FOV. The two outermost points are each $\sim 2\sigma$ below the profiles recovered by Minkasi (and also \emph{XMM}). As is found in many other ground-based single dish SZ experiments, recovery at or beyond the FOV is prone to systematic errors when TODs are processed by subtracting a common-mode (or principle components).  Interestingly, the MIDAS A10 profile shows a higher central pressure, which could be due to point sources degeneracies coupling to gNFW parameter degeneracies.
 
\subsection{Mass Estimates}

We repeat the mass estimations that were done on the Minkasi-derived pressure profiles (Sections~\ref{sec:mass_derivations} and \ref{sec:masses} for the MIDAS branch and tabulate the results in Table~\ref{tbl:m2_mass_MIDAS}. Within the MIDAS branch, $M_{\text{HE}}$ (both $M_{2500}$ and $M_{500}$) are clearly lower than their other respective mass estimates; this is due to the poor pressure profile recovery at and beyond the radial FOV within the MIDAS branch. It is clear that the $M_{500}$ is not physical given the $M_{2500}$ hydrostatic equilibrium. Indeed, the hydrostatic mass curve decreases beyond $\sim120$\arcsec, thus it is not that the $M_{500}$ is found incorrectly, vis-a-vis self-consistency, but this reiterates that the pressure profile is erroneously biased low at and beyond the radial FOV.

Knowing the difficulties MIDAS has recovering the pressure profile beyond MUSTANG-2's radial FOV, we additionally estimate $M_{2500}$, where $R_{2500}$ lies close to the radial FOV (see Figure~\ref{fig:midas_pp}). Unfortunately, we find that the NP model recovers significantly lower masses, with the exception of $M_{\text{VT}}$. However, the masses derived from the A10 pressure profile (gNFW) fits are in in good agreement with those from Minkasi. We conclude that the MIDAS processing performs equally well as Minkasi when determining pressure interior to our FOV, but that some functional fit (e.g. gNFW) should be used with MIDAS at larger radii.

\section{appendix Radio Sources in Zwicky 3146}
\label{sec:appendix_radio}
        All radio sources of concern for MUSTANG-2 are tabulated in Table~\ref{tbl:zwicky_radio}. The six sources with 1.4 GHz data, from FIRST \citep{becker1994} are fit alongside some cluster model. The remaining two sources (S2 and S3) are fit in the MUSTANG-2 residual map.
    
\begin{deluxetable*}{cccccccccccc}
\tabletypesize{\scriptsize}
\tablecolumns{12}
\tablewidth{0pt}
\tablecaption{Zwicky 3146 Radio Sources \label{tbl:zwicky_radio}}
\tablehead{ 
  \colhead{ID} & \colhead{R.A.} & \colhead{Dec.} & \colhead{$S_{0.61}$} & \colhead{$S_{1.4}$} & \colhead{$S_{4.9}$} & \colhead{$S_{8.5}$} & \colhead{${}^c S_{28.5}$} &  \colhead{$S_{90}$} & \colhead {$S_{600}$} & \colhead {$S_{850}$} & \colhead {$S_{1200}$} \\
  J2000 & J2000 & (mJy) & (mJy) & (mJy) & (mJy) & (mJy) & (mJy) & (mJy) & (mJy) & (mJy) 
} 
\startdata
	 S1 & $10^{h}23^{m}39^{s}$.66 & +$04^{\circ}11^{\prime}10^{\prime\prime}$.8 & $7.0\pm0.4$ & $2.04\pm0.15$  & ${}^{a} 1.42\pm0.07$  & ${}^{a} 0.98\pm0.03$  & $0.41\pm0.07$ & $0.191\pm0.022$ & -- & $29 \pm 5$ & $95 \pm 8$ \\
	 S2 & $10^{h}23^{m}38^{s}$.7   & +$04^{\circ}11^{\prime}05^{\prime\prime}$    & --    & --    & ${}^{a} 0.31\pm0.02$  & ${}^{a} 0.37\pm0.02$  & --& $0.047 \pm 0.008$ & -- & -- & --  \\
	 S3 & $10^{h}23^{m}42^{s}$.3  & +$04^{\circ}11^{\prime}3^{\prime\prime}$  & --    & --    & -- & --  & -- & $0.035 \pm 0.008$             & $34 \pm 5$ & $35\pm5$ & $31\pm12$ \\
     S4 & $10^{h}23^{m}44^{s}$.81 & +$04^{\circ}10^{\prime}36^{\prime\prime}$.7 & --    & $32.52\pm0.15$ & -- & --  & ${}^{d} 2.83\pm0.31$ & $0.938\pm0.016$ & -- & --  \\
     S5 & $10^{h}23^{m}45^{s}$.26 & +$04^{\circ}10^{\prime}42^{\prime\prime}$.8 & --    & $56.69\pm0.15$ & -- & ${}^{b} 41\pm11$  & ${}^{d} 2.78\pm0.31$ & $0.741\pm0.015$ & -- & --  & --  \\
     S6 & $10^{h}23^{m}37^{s}$.51 & +$04^{\circ}09^{\prime}13^{\prime\prime}$.2 & --    & $15.06\pm0.15$ & -- & --  & ${}^{e} 1.19\pm0.16$ & $0.494\pm0.019$ & -- & -- & --   \\
     S7 & $10^{h}23^{m}36^{s}$.86 & +$04^{\circ}08^{\prime}59^{\prime\prime}$.0 & --    & $12.07\pm0.15$ & -- & --  & ${}^{e} 0.90\pm0.16$ & $0.365\pm0.020$ & -- & -- & --   \\
     S8 & $10^{h}23^{m}45^{s}$.27 & +$04^{\circ}11^{\prime}41^{\prime\prime}$.0 & --    & $6.06\pm0.15$  & -- & --  & $0.85\pm0.1$ & $0.176\pm0.015$ & -- & -- & --   \\
\enddata
\tablecomments{Radio sources in Zwicky 3146. The total (integrated) flux density for MUSTANG-2 sources is roughly 2 mJy.  $^{a}$ From \citet{giacintucci2014} with VLA in C configuration. ${}^{b}$ From \citet{cooray1998a}; where FIRST saw two sources nearby the coordinates 10:23:45, +04:10:40, \citet{cooray1998a} categorized it as one source. ${}^{c}$ All 28.5 GHz flux densities transcribed here come from OVRO/BIMA \citep{lancaster2011}; ${}^{d}$ and ${}^{e}$ had a single value reported, corresponding to the blending (sum) of the sources in OVRO/BIMA.}
\end{deluxetable*}



\begin{figure}
  \begin{center}
     \includegraphics[width=0.45\textwidth]{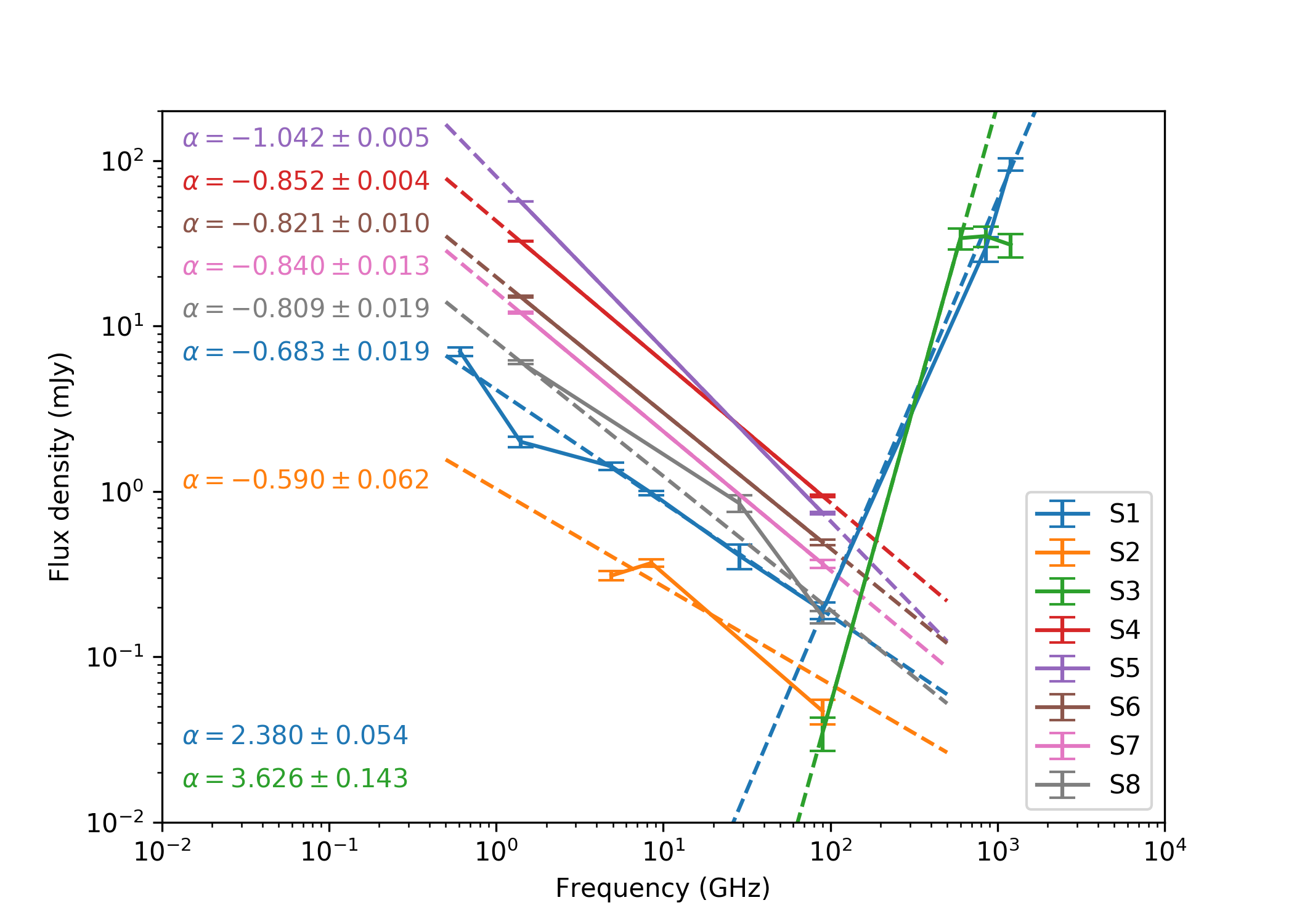}
  \end{center}
  \caption{Spectral Energy Distributions (SEDs) for all point sources tabulated in Table~\ref{tbl:zwicky_radio}. S4, S5, S6, and S7 have interpolated/fitted points shown with circles and errors with no caps.}
  \label{fig:zw3146_SEDs}
\end{figure}

    Here, we briefly investigate the nature of the radio sources from data available online or in the literature. A plot of the spectral energy distributions (SEDs) of the sources is found in Figure~\ref{fig:zw3146_SEDs}. For the sources with low frequency data ($\nu \leq 90$ GHz), we calculate (single) power laws up to 90 GHz, and find that they are good fits except for S2.
    We note that S2 may not be in the cluster as \citet{giacintucci2014} report the photometric redshift from SDSS for S2 as $z_{phot} = 0.34$, which would put it behind the cluster. 
    
    Sources S4 and S5 appear as a slightly extended source in MUSTANG-2. The same is true for S6 and S7. The separation is provided by FIRST, thus we follow it here. \citet{lancaster2011} tabulate two sources which correspond to (S4,S5) and (S6,S7). That is, the OVRO/BIMA data used did not appear to sufficiently resolve these pairs of sources. We calculate a power law from the FIRST and MUSTANG-2 data and find excellent agreement with the sum of their expected flux densities at 28.5 GHz and those reported (for their respective sums) in \citet{lancaster2011}. In Table~\ref{tbl:zwicky_radio}, we have divided the flux density the pair constituents proportional with their expected flux densities. These are also shown with circles in Figure~\ref{fig:zw3146_SEDs}.

    S1, the cluster Brightest Central Galaxy (BCG), and S3 have two and three measured flux densities in publicly available Herschel SPIRE data. We obtain photometry of point sources from the level 2 maps with \lstinline[language=python]!photutils!\footnote{\url{https://photutils.readthedocs.io/en/stable/index.html}}, and incorporate the 5.5\% calibration error noted in the Quick-Start Guide\footnote{\url{https://www.cosmos.esa.int/documents/12133/1035800/QUICK-START+GUIDE+TO+HERSCHEL-SPIRE}}. We do not see a peak in the BCG SED, but we do see a peak for S3. 
    
    For S3, we should thus be able to fit a modified blackbody curve:
    \begin{equation}
        I_{\nu} = I_0 \frac{\nu^{3+\beta}}{e^{h \nu / k_B T} - 1},
    \end{equation}
    where $I_0$ is the spectral irradiance normalization, $\nu$ is the frequency, $h$ is the Planck constant, $k_B$ is the Boltzmann constant, and $T$ is the (redshifted) temperature of the emitting medium, and $\beta$ is the modification to the standard blackbody curve due to (dust) opacity. We find $\beta = 2.7 \pm 0.4$ and $T = 7.2 \pm 1.1$, with $\chi^2 = 2.31$ and 1 degree of freedom, the probability to exceed ($\chi^2$) due to noise alone is 0.128. This value of $\beta$ is higher than perhaps expected \citep{draine1984}, although $\beta > 2$ has been found \citep[e.g][]{kato2018}, and even $\beta > 3$ have been found \citep[e.g. references within][]{shetty2009}. As in \citet{kato2018}, source blending may also affect our Herschel photometry.
    
\end{document}